\documentclass[12pt,a4paper]{article}
\usepackage{comment}
\usepackage[
backend=biber,
style=numeric-comp,
sorting=none
]{biblatex}
\DeclareUnicodeCharacter{2212}{-}

\usepackage[utf8]{inputenc}
\usepackage{amsmath}
\usepackage{amsfonts} 
\usepackage{mathtools} 
\usepackage{amssymb} 
\usepackage{subfig} 
\usepackage{graphicx}
\usepackage{xcolor}
\usepackage{mathrsfs}
\usepackage{hyperref}
\usepackage{multirow}
\usepackage[normalem]{ulem}

\usepackage[page, toc]{appendix}

\begin{document}
{
\renewcommand{\refeq}[1]{eq.\,\eqref{#1}} 
\newcommand{\refEq}[1]{Eq.\,\eqref{#1}} 
\newcommand{\refeqs}[1]{eqs.\,\eqref{#1}} 
\newcommand{\refEqs}[1]{Eqs.\,\eqref{#1}} 
\newcommand{\abs}[1]{|#1|} 
\newcommand{\Abs}[1]{\left|#1\right|} 
\newcommand{\re}[1]{\text{Re}\left(#1\right)}
\newcommand{\im}[1]{\text{Im}\left(#1\right)}
\newcommand{\Hc}{\text{H.c.}}
\newcommand{\id}{\mathbf{1}}
\newcommand{\TR}[1]{\text{Tr}\left\{#1\right\}}
\newcommand{\BR}[1]{\text{Br}\left(#1\right)}
\newcommand{\PL}{P_L}\newcommand{\PR}{P_R}
\newcommand{\Zn}[1]{\mathbb{Z}_{#1}}
\newcommand{\ZZ}{\Zn{2}}
\newcommand{\cb}{c_\beta}
\renewcommand{\sb}{s_\beta}
\newcommand{\cbb}{c_{2\beta}}
\newcommand{\sbb}{s_{2\beta}}
\newcommand{\cba}{c_{\alpha\beta}}
\newcommand{\sba}{s_{\alpha\beta}}
\newcommand{\tb}{t_\beta}
\newcommand{\tbb}{t_{2\beta}}
\newcommand{\tbinv}{\tb^{-1}}
\newcommand{\tti}{\tb+\tbinv}
\newcommand{\VEV}[1]{\langle #1 \rangle}
\newcommand{\vev}[1]{v_{#1}}
\newcommand{\ROTmat}{\mathcal R}
\newcommand{\ROTmatT}{\ROTmat^T}
\newcommand{\ROTmatinv}{\ROTmat^{-1}}
\newcommand{\ROT}[1]{\ROTmat_{#1}}
\newcommand{\HbROT}{\mathcal R_{\beta}^{\phantom{T}}}
\newcommand{\HbROTt}{\mathcal R_{\beta}^T}
\newcommand{\HbROTinv}{\mathcal R_{\beta}^{-1}}
\newcommand{\NSROT}{\mathcal R_{[3]}(\vec{\alpha})}
\newcommand{\NSROTt}{\mathcal R_{[3]}(\vec{\alpha})^T}
\newcommand{\NSROTinv}{\mathcal R_{[3]}(\vec{\alpha})^{-1}}
\newcommand{\SMHD}{\Phi}
\newcommand{\SMHDd}{\Phi^\dagger}
\newcommand{\HD}[1]{\Phi_{#1}^{\phantom{\dagger}}}
\newcommand{\HDd}[1]{\Phi_{#1}^\dagger}
\newcommand{\HDC}[1]{\tilde\Phi_{#1}^{\phantom{\dagger}}}
\newcommand{\HDc}[1]{\Phi_{#1}^{\phantom{\dagger}\!\!\!\ast}}
\newcommand{\HHD}[1]{H_{#1}^{\phantom{\dagger}}}
\newcommand{\HHDd}[1]{H_{#1}^\dagger}
\newcommand{\HHDC}[1]{\tilde H_{#1}^{\phantom{\dagger}}}
\newcommand{\HHDc}[1]{H_{#1}^{\phantom{\dagger}\!\!\!\ast}}
\newcommand{\nHH}{{H}^0}
\newcommand{\nHR}{{R}^0}
\newcommand{\nHI}{{I}^0}
\newcommand{\nh}{\mathrm{h}}
\newcommand{\nhSM}{\mathrm{h_{SM}}}
\newcommand{\nH}{\mathrm{H}}
\newcommand{\nA}{\mathrm{A}}
\newcommand{\nS}{\mathrm{S}}
\newcommand{\cH}{\mathrm{H}^\pm}
\newcommand{\cHm}{\mathrm{H}^-}
\newcommand{\cHp}{\mathrm{H}^+}
\newcommand{\mNSc}{\mathcal M_0^2}
\newcommand{\mNScT}{{\mathcal M_0^{2}}^T}
\newcommand{\mh}{m_{\nh}}
\newcommand{\mhSM}{m_{\nhSM}}
\newcommand{\mH}{m_{\nH}}
\newcommand{\mA}{m_{\nA}}
\newcommand{\mcH}{m_{\cH}}
\newcommand{\mS}{m_{\mathrm{S}}}
\newcommand{\nl}[1]{n_{#1}}
\newcommand{\nrl}[1]{\re{n_{#1}}}
\newcommand{\nrle}{\nrl{e}}\newcommand{\nrlm}{\nrl{\mu}}\newcommand{\nrlt}{\nrl{\tau}}
\newcommand{\noml}[1]{\frac{\nrl{#1}}{m_{#1}}}

\newcommand{\Yd}[1]{\Gamma_{#1}}
\newcommand{\Yu}[1]{\Delta_{#1}}
\newcommand{\Ydc}[1]{\Gamma_{#1}^\ast}
\newcommand{\Yuc}[1]{\Delta_{#1}^\ast}
\newcommand{\Ydd}[1]{\Gamma_{#1}^\dagger}
\newcommand{\Yud}[1]{\Delta_{#1}^\dagger}
 \newcommand{\FQ}{Q}\newcommand{\FL}{L}
\newcommand{\Fu}{u}\newcommand{\Fd}{d}
\newcommand{\Fl}{\ell}\newcommand{\Fn}{\nu}
\newcommand{\ferX}[3]{{#1}_{#2#3}}\newcommand{\ferXb}[3]{\bar #1_{#2#3}}
\newcommand{\dL}[1]{\ferX{\Fd}{L}{#1}}\newcommand{\dLb}[1]{\ferXb{\Fd}{L}{#1}}
\newcommand{\dR}[1]{\ferX{\Fd}{R}{#1}}\newcommand{\dRb}[1]{\ferXb{\Fd}{R}{#1}}
\newcommand{\uL}[1]{\ferX{\Fu}{L}{#1}}\newcommand{\uLb}[1]{\ferXb{\Fu}{L}{#1}}
\newcommand{\uR}[1]{\ferX{\Fu}{R}{#1}}\newcommand{\uRb}[1]{\ferXb{\Fu}{R}{#1}}
\newcommand{\lL}[1]{\ferX{\Fl}{L}{#1}}\newcommand{\lLb}[1]{\ferXb{\Fl}{L}{#1}}
\newcommand{\lR}[1]{\ferX{\Fl}{R}{#1}}\newcommand{\lRb}[1]{\ferXb{\Fl}{R}{#1}}
\newcommand{\nL}[1]{\ferX{\Fn}{L}{#1}}\newcommand{\nLb}[1]{\ferXb{\Fn}{L}{#1}}
\newcommand{\nR}[1]{\ferX{\Fn}{R}{#1}}\newcommand{\nRb}[1]{\ferXb{\Fn}{R}{#1}}

\newcommand{\SD}[1]{\Phi_{#1}^{\phantom{\dagger}}}
\newcommand{\SDd}[1]{\Phi_{#1}^\dagger}
\newcommand{\SDc}[1]{\Phi_{#1}^\ast}
\newcommand{\SDti}[1]{\tilde\Phi_{#1}^{\phantom{\dagger}}}
\newcommand{\bilSD}[2]{\SDd{#1}\SD{#2}}
\newcommand{\pbilSD}[2]{\big(\bilSD{#1}{#2}\big)}

\newcommand{\Hv}{H_1}\newcommand{\Hvd}{H_1^\dagger}\newcommand{\Hvti}{\tilde H_1}
\newcommand{\Ho}{H_2}\newcommand{\Hod}{H_2^\dagger}\newcommand{\Hoti}{\tilde H_2}

\newcommand{\weakferX}[3]{{#1}_{#2#3}^0}\newcommand{\weakferXb}[3]{\bar #1_{#2#3}^0}

\newcommand{\wQL}[1]{\weakferX{\FQ}{L}{#1}}\newcommand{\wQLb}[1]{\weakferXb{\FQ}{L}{#1}}
\newcommand{\wdL}[1]{\weakferX{\Fd}{L}{#1}}\newcommand{\wdLb}[1]{\weakferXb{\Fd}{L}{#1}}
\newcommand{\wdR}[1]{\weakferX{\Fd}{R}{#1}}\newcommand{\wdRb}[1]{\weakferXb{\Fd}{R}{#1}}
\newcommand{\wuL}[1]{\weakferX{\Fu}{L}{#1}}\newcommand{\wuLb}[1]{\weakferXb{\Fu}{L}{#1}}
\newcommand{\wuR}[1]{\weakferX{\Fu}{R}{#1}}\newcommand{\wuRb}[1]{\weakferXb{\Fu}{R}{#1}}
\newcommand{\wLL}[1]{\weakferX{\FL}{L}{#1}}\newcommand{\wLLb}[1]{\weakferXb{\FL}{L}{#1}}
\newcommand{\wlL}[1]{\weakferX{\Fl}{L}{#1}}\newcommand{\wlLb}[1]{\weakferXb{\Fl}{L}{#1}}
\newcommand{\wlR}[1]{\weakferX{\Fl}{R}{#1}}\newcommand{\wlRb}[1]{\weakferXb{\Fl}{R}{#1}}
\newcommand{\wnL}[1]{\weakferX{\Fn}{L}{#1}}\newcommand{\wnLb}[1]{\weakferXb{\Fn}{L}{#1}}
\newcommand{\QL}[1]{\ferX{\FQ}{L}{#1}}\newcommand{\QLb}[1]{\ferXb{\FQ}{L}{#1}}
\newcommand{\LL}[1]{\ferX{\FL}{L}{#1}}\newcommand{\LLb}[1]{\ferXb{\FL}{L}{#1}}

\newcommand{\CKM}{V}\newcommand{\CKMdag}{V^\dagger}
\newcommand{\V}[1]{\CKM^{\phantom{\ast}}_{#1}}
\newcommand{\Vc}[1]{\CKM^{\ast}_{#1}}
\newcommand{\Vd}[1]{\CKM^{\dagger}_{#1}}\newcommand{\Vt}[1]{\CKM^{t}_{#1}}
\newcommand{\PMNS}{U}\newcommand{\PMNSdag}{U^\dagger}
\newcommand{\U}[1]{\PMNS^{\phantom{\ast}}_{#1}}
\newcommand{\Uc}[1]{\PMNS^{\ast}_{#1}}
\newcommand{\Ud}[1]{\PMNS^{\dagger}_{#1}}\newcommand{\Ut}[1]{\PMNS^{t}_{#1}}
\newcommand{\matXF}[2]{{\rm #1}_{#2}}\newcommand{\matXFd}[2]{{\rm #1}_{#2}^\dagger}
\newcommand{\wmatXF}[2]{{\rm #1}_{#2}^{0}}\newcommand{\wmatXFd}[2]{{\rm #1}_{#2}^{0\dagger}}
\newcommand{\basematM}{M}\newcommand{\basematN}{N}
\newcommand{\matNf}[1]{\matXF{\basematN}{#1}}\newcommand{\matNfd}[1]{\matXFd{\basematN}{#1}}
\newcommand{\matMf}[1]{\matXF{\basematM}{#1}}\newcommand{\matMfd}[1]{\matXFd{\basematM}{#1}}
\newcommand{\matND}{\matXF{\basematN}{d}}\newcommand{\matNDd}{\matXFd{\basematN}{d}}
\newcommand{\matNU}{\matXF{\basematN}{u}}\newcommand{\matNUd}{\matXFd{\basematN}{u}}
\newcommand{\matNL}{\matXF{\basematN}{\ell}}\newcommand{\matNLd}{\matXFd{\basematN}{\ell}}
\newcommand{\matNN}{\matXF{\basematN}{\nu}}\newcommand{\matNNd}{\matXFd{\basematN}{\nu}}
\newcommand{\matMD}{\matXF{\basematM}{d}}\newcommand{\matMDd}{\matXFd{\basematM}{d}}
\newcommand{\matMU}{\matXF{\basematM}{u}}\newcommand{\matMUd}{\matXFd{\basematM}{u}}
\newcommand{\matML}{\matXF{\basematM}{\ell}}\newcommand{\matMLd}{\matXFd{\basematM}{\ell}}
\newcommand{\matMN}{\matXF{\basematM}{\nu}}\newcommand{\matMNd}{\matXFd{\basematM}{\nu}}
\newcommand{\wmatNf}{\wmatXF{\basematN}{f}}\newcommand{\wmatNDf}{\wmatXF{\basematN}{f}}
\newcommand{\wmatMf}{\wmatXF{\basematM}{f}}\newcommand{\wmatMDf}{\wmatXF{\basematM}{f}}
\newcommand{\wmatND}{\wmatXF{\basematN}{d}}\newcommand{\wmatNDd}{\wmatXFd{\basematN}{d}}
\newcommand{\wmatNU}{\wmatXF{\basematN}{u}}\newcommand{\wmatNUd}{\wmatXFd{\basematN}{u}}
\newcommand{\wmatNL}{\wmatXF{\basematN}{\ell}}\newcommand{\wmatNLd}{\wmatXFd{\basematN}{\ell}}
\newcommand{\wmatNN}{\wmatXF{\basematN}{\nu}}\newcommand{\wmatNNd}{\wmatXFd{\basematN}{\nu}}
\newcommand{\wmatMD}{\wmatXF{\basematM}{d}}\newcommand{\wmatMDd}{\wmatXFd{\basematM}{d}}
\newcommand{\wmatMU}{\wmatXF{\basematM}{u}}\newcommand{\wmatMUd}{\wmatXFd{\basematM}{u}}
\newcommand{\wmatML}{\wmatXF{\basematM}{\ell}}\newcommand{\wmatMLd}{\wmatXFd{\basematM}{\ell}}
\newcommand{\wmatMN}{\wmatXF{\basematM}{\nu}}\newcommand{\wmatMNd}{\wmatXFd{\basematM}{\nu}}

\newcommand{\UfX}[2]{\mathcal U_{#1_{#2}}}\newcommand{\UfXd}[2]{\mathcal U_{#1_{#2}}^\dagger}
\newcommand{\UuL}{\UfX{u}{L}}\newcommand{\UuLd}{\UfXd{u}{L}}\newcommand{\UuR}{\UfX{u}{R}}\newcommand{\UuRd}{\UfXd{u}{R}}
\newcommand{\UdL}{\UfX{d}{L}}\newcommand{\UdLd}{\UfXd{d}{L}}\newcommand{\UdR}{\UfX{d}{R}}\newcommand{\UdRd}{\UfXd{d}{R}}
\newcommand{\UlL}{\UfX{\ell}{L}}\newcommand{\UlLd}{\UfXd{\ell}{L}}\newcommand{\UlR}{\UfX{\ell}{R}}\newcommand{\UlRd}{\UfXd{\ell}{R}}
\newcommand{\UnL}{\UfX{\nu}{L}}\newcommand{\UnLd}{\UfXd{\nu}{L}}\newcommand{\UnR}{\UfX{\nu}{R}}\newcommand{\UnRd}{\UfXd{\nu}{R}}

\newcommand{\matYukF}[2]{Y_{#1#2}}\newcommand{\matYukFd}[2]{Y_{#1#2}^\dagger}\newcommand{\matYukFt}[2]{Y_{#1#2}^t}\newcommand{\matYukFc}[2]{Y_{#1#2}^\ast}
\newcommand{\YukF}[3]{(Y_{#1#2})_{#3}}\newcommand{\YukFd}[3]{(Y_{#1#2}^\dagger)_{#3}}\newcommand{\YukFt}[3]{(Y_{#1#2}^t)_{#3}}\newcommand{\YukFc}[3]{(Y_{#1#2}^\ast)_{#3}}
\newcommand{\matYukU}[1]{\matYukF{u}{#1}}\newcommand{\matYukUd}[1]{\matYukFd{u}{#1}}\newcommand{\matYukUt}[1]{\matYukFt{u}{#1}}\newcommand{\matYukUc}[1]{\matYukFc{u}{#1}}
\newcommand{\YukU}[2]{\YukF{u}{#1}{#2}}\newcommand{\YukUd}[2]{\YukFd{u}{#1}{#2}}\newcommand{\YukUt}[2]{\YukFt{u}{#1}{#2}}\newcommand{\YukUc}[2]{\YukFc{u}{#1}{#2}}
\newcommand{\matYukD}[1]{\matYukF{d}{#1}}\newcommand{\matYukDd}[1]{\matYukFd{d}{#1}}\newcommand{\matYukDt}[1]{\matYukFt{d}{#1}}\newcommand{\matYukDc}[1]{\matYukFc{d}{#1}}
\newcommand{\YukD}[2]{\YukF{d}{#1}{#2}}\newcommand{\YukDd}[2]{\YukFd{d}{#1}{#2}}\newcommand{\YukDt}[2]{\YukFt{d}{#1}{#2}}\newcommand{\YukDc}[2]{\YukFc{d}{#1}{#2}}
\newcommand{\matYukL}[1]{\matYukF{\ell}{#1}}\newcommand{\matYukLd}[1]{\matYukFd{\ell}{#1}}\newcommand{\matYukLt}[1]{\matYukFt{\ell}{#1}}\newcommand{\matYukLc}[1]{\matYukFc{\ell}{#1}}
\newcommand{\YukL}[2]{\YukF{\ell}{#1}{#2}}\newcommand{\YukLd}[2]{\YukFd{\ell}{#1}{#2}}\newcommand{\YukLt}[2]{\YukFt{\ell}{#1}{#2}}\newcommand{\YukLc}[2]{\YukFc{\ell}{#1}{#2}}
\newcommand{\matYukN}[1]{\matYukF{\nu}{#1}}\newcommand{\matYukNd}[1]{\matYukFd{\nu}{#1}}\newcommand{\matYukNt}[1]{\matYukFt{\nu}{#1}}\newcommand{\matYukNc}[1]{\matYukFc{\nu}{#1}}
\newcommand{\YukN}[2]{\YukF{\nu}{#1}{#2}}\newcommand{\YukNd}[2]{\YukFd{\nu}{#1}{#2}}\newcommand{\YukNt}[2]{\YukFt{\nu}{#1}{#2}}\newcommand{\YukNc}[2]{\YukFc{\nu}{#1}{#2}}

\newcommand{\glFCno}{g$\ell$FC}
\newcommand{\glFC}[1]{#1-\glFCno}
\newcommand{\solA}{[A]}
\newcommand{\solB}{[B]}
\newcommand{\solBpm}{[B$_\pm$]}
\newcommand{\solBp}{[B$_+$]}
\newcommand{\solBm}{[B$_-$]}


 \hfill\begin{minipage}[r]{0.3\textwidth}\begin{flushright}  IFIC/23-05 \end{flushright} \end{minipage}

\begin{center}

\vspace{0.50cm}
{\large\bf {New Physics hints from $\tau$ scalar interactions and $(g-2)_{e,\mu}$}}

\vspace{0.50cm}
Francisco J. Botella $^{a,}$\footnote{\texttt{Francisco.J.Botella@uv.es}}, 
Fernando Cornet-Gomez $^{b,}$\footnote{\texttt{Fernando.CornetGomez@case.edu}}, Carlos Miró $^{a,}$\footnote{\texttt{Carlos.Miro@uv.es}},
Miguel Nebot $^{a,}$\footnote{\texttt{Miguel.Nebot@uv.es}}
\end{center}
\vspace{0.50cm}
\begin{flushleft}
\emph{$^a$ Departament de F\' \i sica Te\`orica and IFIC, Universitat de Val\`encia-CSIC,\\ \quad E-46100, Burjassot, Spain.} \\
\emph{$^b$ Physics Department and Center for Education and Research in Cosmology and Astrophysics (CERCA), Case Western Reserve University, Cleveland, OH 44106, USA.}
\end{flushleft}
\vspace{0.5cm}
\date{\today}
\begin{abstract}
We consider a flavour conserving two Higgs doublet model that consists of a type I (or X) quark sector and a generalized lepton sector where the Yukawa couplings of the charged leptons to the new scalars are not proportional to the lepton masses. The model, previously proposed to solve both muon and electron $g-2$ anomalies simultaneously, is also capable to accommodate the ATLAS excess in $pp \rightarrow S \rightarrow \tau^{+}\tau^{-}$ with gluon-gluon fusion production in the invariant mass range [0.2; 0.6] TeV, including all relevant low and high energy constraints. The excess is reproduced taking into account the new contributions from the scalar H, the pseudoscalar A, or both. In particular, detailed numerical analyses favoured the solution with a significant hierarchy among the vevs of the two Higgs doublets, $\tb \sim 10$, and light neutral scalars satisfying $\mA > \mH$ with sizable couplings to $\tau$ leptons. In this region of the parameter space, the muon $g-2$ anomaly receives one and two loop (Barr-Zee) contributions of similar size, while the electron anomaly is explained at two loops. An analogous ATLAS excess in $b$-associated production and the CMS excess in ditop production are also studied. Further New Physics prospects concerning the anomalous magnetic moment of the $\tau$ lepton and the implications of the CDF $M_W$ measurement on the final results are discussed.
\end{abstract}
\section{Introduction\label{SEC:Intro}}
%
The experimental determination of the anomalous magnetic moments $a_\ell=\frac{g_\ell-2}{2}$, $\ell=\{e,\mu\}$, of both the electron and the muon ($g_\ell$ are the gyromagnetic ratios) appear to deviate from the Standard Model (SM) expectations. The most recent result from the \emph{Muon g-2} collaboration \cite{Muong-2:2021ojo,Muong-2:2021vma} is
\begin{equation}\label{eq:muon:anomaly}
\delta a_\mu^{\rm Exp}=a_\mu^{\rm Exp}-a_\mu^{\rm SM}=(2.5\pm 0.6)\times 10^{-9},
\end{equation}
where $a_\mu^{\rm Exp}$ is the measured value and $a_\mu^{\rm SM}$ is the SM prediction \cite{Aoyama:2020ynm,Aoyama:2012wk,Aoyama:2019ryr,Czarnecki:2002nt,Gnendiger:2013pva,Davier:2017zfy,Keshavarzi:2018mgv,Colangelo:2018mtw,Hoferichter:2019mqg,Davier:2019can,Keshavarzi:2019abf,Kurz:2014wya,Melnikov:2003xd,Masjuan:2017tvw,Colangelo:2017fiz,Hoferichter:2018kwz,Gerardin:2019vio,Bijnens:2019ghy,Colangelo:2019uex,Blum:2019ugy,Colangelo:2014qya}.
For the electron anomalous magnetic moment, precise determinations of the fine structure constant using atomic recoils \cite{Hanneke:2008tm,Parker:2018vye,Morel:2020dww,Tiesinga:2021myr} together with impressive calculations of the SM expectation $a_e^{\rm SM}$ \cite{Aoyama:2012wj,Aoyama:2012wk,Laporta:2017okg,Aoyama:2017uqe,Volkov:2019phy} yield 
\begin{equation}\label{eq:electron:anomaly}
\delta a_e^{\rm Exp,Cs}=-(8.7\pm 3.6)\times 10^{-13},\qquad 
\delta a_e^{\rm Exp,Rb}=(4.8\pm 3.0)\times 10^{-13}.
\end{equation}
The possibility to explain both $\delta a_\mu^{\rm Exp}$ and $\delta a_e^{\rm Exp}$ deviations simultaneously has been explored in different scenarios \cite{Haba:2020gkr,Jana:2020pxx,Dutta:2020scq,DelleRose:2020oaa,Han:2021gfu,Li:2020dbg,Liu:2018xkx,Endo:2020mev,Arbelaez:2020rbq,Jana:2020joi,Endo:2019bcj,Bigaran:2020jil,Calibbi:2020emz,Bodas:2021fsy,Borah:2021khc,Li:2021wzv,Crivellin:2018qmi,Bauer:2019gfk,Calibbi:2020emz,Jana:2020joi,Fajfer:2021cxa,Dcruz:2022dao,Chen:2023eof,Hiller:2019mou}. In particular, in references \cite{Botella:2020xzf} and \cite{Botella:2022rte}, they have been addressed in the context of two Higgs doublet models (2HDMs) with a particular flavour structure where tree level scalar neutral flavour changing couplings are absent. Specifically, the Yukawa quark sector is analogous to a $\ZZ$-shaped 2HDM of type I, while in the lepton sector the most general flavour conserving scenario is considered \cite{Penuelas:2017ikk,Botella:2018gzy}.

With massless neutrinos and no CP violation in the Yukawa and scalar sectors --the latter a simplifying assumption that guarantees vanishing electric dipole moments \cite{ACME:2018yjb}--, one new parameter is introduced for each charged lepton: $\nl{e}$, $\nl{\mu}$ and $\nl{\tau}$. The New Physics (NP) contributions involving the new scalars that can explain $\delta a_e^{\rm Exp}$ and $\delta a_\mu^{\rm Exp}$ necessarily involve sizable $\nl{e}$ and $\nl{\mu}$. Although in principle $\nl{\tau}\neq 0$ is not necessary, there are regions of parameter space with a sizable $\nl{\tau}$. In this context, the NP hints that the ATLAS collaboration presented in \cite{ATLAS:2020zms} regarding $pp\to S\to \tau^+\tau^-$ are particularly appealing: they certainly require new neutral scalars with masses in the $[0.2;0.6]$ TeV range which couple significantly to $\tau$ leptons. In this work we analyse how the scenario previously considered to explain $\delta a_e^{\rm Exp}$ and $\delta a_\mu^{\rm Exp}$ can also explain the ATLAS excess in a specific region of parameter space. Furthermore, we also explore the possibility to accommodate the recent CDF measurement of $M_W$ \cite{CDF:2022hxs}, which deviates from the SM expectations derived from electroweak precision data, through a modification of the oblique parameters $S$ and $T$.

The paper has the following organization. In section \ref{SEC:Model} we introduce the main aspects of the model. In section \ref{SEC:Paper1} we discuss how the model can account for $\delta a_e^{\rm Exp}$ and $\delta a_\mu^{\rm Exp}$ after relevant constraints are imposed. In section \ref{SEC:ATLAS} we comment in detail the ATLAS excess in \cite{ATLAS:2020zms}. Section \ref{SEC:Results} is devoted to discuss results of the analyses which incorporate an explanation of the ATLAS excess within this model. Further potential NP hints are addressed in section \ref{SEC:NP}. Finally, we disclose our conclusions in section \ref{SEC:Conclusions}. We relegate to an appendix some details concerning the construction of the proposed model.

\section{The model} \label{SEC:Model}
The 2HDM Yukawa Lagrangian in the \textit{Higgs basis} \cite{GEORGI197995,PhysRevD.19.945,PhysRevD.51.3870}, where only one scalar doublet acquires a vacuum expectation value (vev), takes the form
\begin{equation}\label{eq:2HDM:YukLag:HiggsBasis:Diagonal}%
\begin{split}
\mathscr L_{\rm Y} =&-\frac{\sqrt{2}}{v}\QLb{}\left(\Hv\matMD+\Ho\matND\right)\dR{} -\frac{\sqrt{2}}{v}\QLb{}\left(\Hvti\matMU+\Hoti\matNU\right)\uR{}\\
&-\frac{\sqrt{2}}{v}\LLb{}\left(\Hv\matML+\Ho\matNL\right)\lR{}+\Hc \, , 
\end{split}
\end{equation}
with $v^2 = v_1^2 + v_2^2 = (\sqrt{2} G_F)^{-1} \simeq (246\ \mathrm{GeV})^2$. The fermion fields in \refeq{eq:2HDM:YukLag:HiggsBasis:Diagonal} have already been rotated into their mass eigenstates, so that the matrices $\matMf{f}$ ($f = d, u, \ell$) represent diagonal fermion mass matrices. On the other hand, the new flavour structures $\matNf{f}$ satisfy
\begin{equation}\label{eq:Nmatrices}
\matNU=\tbinv\matMU,\quad \matND=\tbinv\matMD,\quad \matNL=\text{diag}(\nl{e},\nl{\mu},\nl{\tau})\, , 
\end{equation}
with $\tb\equiv\tan\beta=\vev{2}/\vev{1}$ being the ratio of the vevs of the scalar doublets in an arbitrary weak basis.\footnote{Henceforth, $\tbinv \equiv \cot{\beta}$.} It is clear that our \glFC{I} model consists of a type I (or type X) quark sector and a general Flavour Conserving lepton sector. The $\matNf{\ell}$ matrices in the lepton sector are diagonal, arbitrary and one loop stable under renormalization group evolution (RGE) \cite{Botella:2018gzy}, that is, although the different Yukawa couplings (and $\tb$) run, \refeq{eq:Nmatrices}, with their evolved values, is fulfilled. In particular, one should note that the new Yukawa couplings of the charged leptons are decoupled from lepton mass proportionality, meaning that $n_e/m_e = n_\mu/m_\mu = n_\tau/m_\tau$ does not hold. In this sense, it is precisely the independence of the different couplings $n_\ell$ that will allow us to explain the $(g-2)_\ell$ anomalies simultaneously. Furthermore, we assume that the new lepton Yukawa couplings are real, i.e. $\text{Im}(\nl{\ell})=0$, to avoid dangerous contributions to electric dipole moments highly constrained by experiments, e.g., $|d_e| < 1.1 \times 10^{−29}\ \text{e}\cdot\text{cm}$ \cite{ACME:2018yjb}.

The physical scalar-fermion interactions are obtained once the neutral scalars are rotated into their mass eigenstates, which are determined by the scalar potential. On that respect, we consider a CP conserving scalar sector shaped by a $\ZZ$ symmetry that is softly broken. 

All in all, the Yukawa couplings of neutral scalars read
\begin{equation}\label{eq:neutral:couplings}
\begin{split}
\mathscr{L_{\rm neutral}} =& -\frac{m_{u_j}}{v} \left(\sba + \cba \tbinv \right) \nh\, \bar{u}_j u_j -\frac{m_{d_j}}{v} \left(\sba + \cba \tbinv \right) \nh\, \bar{d}_j d_j \\
&-\frac{m_{\ell_j}}{v} \left(\sba + \cba \frac{\nrl{\ell_j}}{m_{\ell_j}} \right)  \nh\, \bar{\ell}_j \ell_j \\
& -\frac{m_{u_j}}{v} \left(-\cba + \sba \tbinv \right) \nH\, \bar{u}_j u_j -\frac{m_{d_j}}{v} \left(-\cba + \sba \tbinv \right) \nH\, \bar{d}_j d_j \\
&-\frac{m_{\ell_j}}{v} \left(-\cba + \sba \frac{\nrl{\ell_j}}{m_{\ell_j}} \right)  \nH\, \bar{\ell}_j \ell_j \\
&+i\frac{m_{u_j}}{v} \tbinv \nA\, \bar{u}_j \gamma_5 u_j -i\frac{m_{d_j}}{v} \tbinv \nA\, \bar{d}_j \gamma_5 d_j -i\frac{\nrl{\ell_j}}{v} \nA\, \bar{\ell}_j \gamma_5 \ell_j\, , \\
\end{split}
\end{equation}
which are flavour conserving, and those corresponding to charged scalars are
\begin{equation}\label{eq:charged:couplings}
\begin{split}
\mathscr{L_{\rm charged}} =&\ \frac{\cHm}{\sqrt{2}v}\, \bar{d}_i\, V_{ji}^{*} \tbinv \left[(m_{u_j}-m_{d_i}) + (m_{u_j}+m_{d_i}) \gamma_5\right] u_j \\
&+ \frac{\cHp}{\sqrt{2}v}\, \bar{u}_j\, V_{ji} \tbinv \left[(m_{u_j}-m_{d_i}) - (m_{u_j}+m_{d_i}) \gamma_5\right] d_i \\
&- \frac{\cHm}{\sqrt{2}v}\, \bar{\ell}_j\, \nrl{\ell_j} \left(1-\gamma_5\right) \nu_j - \frac{\cHp}{\sqrt{2}v}\, \bar{\nu}_j\, \nrl{\ell_j} \left(1+\gamma_5\right) \ell_j\, , \\
\end{split}
\end{equation}
with $i,j = 1,2,3$ summing over generations. In our notation, $\sba \equiv \sin(\alpha + \beta)$ and $\cba \equiv \cos(\alpha + \beta)$. Note that in the \textit{alignment limit} where $\sba\to 1$, the field $\nh$ couples to fermions as the SM Higgs. For further details on the construction of the model, see appendix \ref{appendix:Model}.
\section{Explaining $(g-2)_{e,\mu}$ with the \glFC{I} model\label{SEC:Paper1}}
%
Before discussing the ATLAS excess in \cite{ATLAS:2020zms}, it is convenient to revisit the main results presented in \cite{Botella:2022rte} within our \glFC{I} framework. In \cite{Botella:2022rte}, the possibility to reproduce both $(g-2)_{e,\mu}$ lepton anomalies was successfully addressed, including the different scenarios one can consider for the electron anomaly, related to the cesium (Cs) or the rubidium (Rb) atomic recoil measurements of the fine structure constant. On that respect, since the Cs case appeared to be a more challenging scenario, we will only recall the results that involve $\delta a_\mu^{\rm{Exp}}$ and $\delta a_e^{\rm{Exp,Cs}}$ in \refeqs{eq:muon:anomaly} and \eqref{eq:electron:anomaly}, respectively.

In the \glFC{I} model, both one loop and two loop Barr-Zee contributions can play a relevant role to solve the $\delta a_\ell$ anomalies. On the one hand, the dominant one loop diagrams are mediated by the neutral scalars $S = \{\nH,\nA\}$, suppressed by $m_S^2$ and proportional to $|n_\ell|^2$. On the other hand, the dominant two loop Barr-Zee contributions to $\delta a_\ell$ are linear in $\nrle$ or $\nrlm$ and include a global negative sign. Among them, those containing a top quark or a tau lepton in the closed fermion loop can be dominant, depending on the specific region of parameter space that is considered, since there is a $\tbinv$ suppression factor for top quarks and a $\nrlt$ enhancement for tau leptons.

The ultimate answer on how to accommodate the $\delta a_{e,\mu}$ anomalies is provided by a detailed numerical analysis including in addition all relevant low and high energy constraints. We briefly recall these constraints.
\begin{itemize}
    \item Perturbativity and perturbative unitarity of high energy $2\to 2$ scattering constraints on quartic couplings, together with boundedness from below of the scalar potential.
    \item Oblique parameters $S$ and $T$ from electroweak precision data. The inclusion of the CDF measurement of the mass of the $W$ boson \cite{CDF:2022hxs} is discussed in section \ref{SEC:NP}.
    \item Higgs signal strengths for the standard combinations of production mechanism and decay mode; they impose, essentially, alignment in the scalar sector.
    \item $\cH$ contributions to tree level processes: lepton flavour universality constraints in leptonic and semileptonic decays.
    \item $\cH$ contributions to one loop transitions like $b\to s\gamma$ and $B_q$--$\bar B_q$ mixings.
    \item $e^+e^-\to S\to\mu^+\mu^-,\tau^+\tau^-$ cross sections from LEP2.
    \item Perturbativity of the Yukawa couplings.
\end{itemize}
As a summary of the main results obtained in \cite{Botella:2022rte} with such numerical analyses, we show and comment some figures that illustrate the allowed regions in the parameter space of the model.\footnote{As discussed in appendix \ref{appendix:Model}, one can choose the parameters defining the model to be the new couplings $\nl{\ell}$, the new scalar masses $\mH$, $\mA$, $\mcH$, and also, from the scalar sector, $\tb$, $\mu_{12}^2$ and the mixing $\cba$.} These allowed regions correspond to the contours in $\Delta \chi^2 =\chi^2-\chi^2_\mathrm{Min}$ associated, darker to lighter, to 1, 2 and 3$\sigma$ regions (68.2\%, 95.4\% and 99.7\% C.L., respectively) in a 2D-$\Delta\chi^2$ distribution. For additional details on the implementation of all the observables included in the $\chi^2$ function we refer to \cite{Botella:2022rte}.

Figure \ref{sfig:mH:tb:Cs250} shows $\mH$ vs. $\tb$: there is a significant hierarchy in the vevs of the two Higgs doublets, that is $\tb \gg 1$, for scalar masses below 1 TeV, while smaller values of $\tb$, namely $\tb \sim 1$, are related to large scalar masses. The latter region requires new scalars to be approximately degenerate, as can be seen in figure \ref{sfig:mA:mH:Cs250}, where $\mA$ vs. $\mH$ is presented. In the former region one needs, instead, $\mA > \mH$ in order to satisfy the constraints imposed by LHC direct searches of new scalars.

To characterize the behavior of the lepton couplings, figure \ref{sfig:nmu:mH:Cs250} shows $\nrlm$ vs. $\mH$. For light scalar masses, namely $\mH < 1$ TeV, the muon anomaly is explained at one loop through the $\nH$ mediated contribution, leading to the approximate relation $|n_\mu| \sim \mH/4$, where $\nrlm$ can take both signs. For scalar masses larger than 1 TeV, because of the perturbativity requirement on the Yukawa couplings $\nl{\ell}$, two loop contributions are needed to reproduce the muon anomaly. In particular, contributions with virtual top quarks are the dominant ones, since there is no $\tbinv$ suppression in that regime of large scalar masses. Furthermore, it is clear that in this region the sign of $\nrlm$ is fixed to be negative, as one could have expected taking into account that the two loop contribution to $\delta a_\mu$ is linear in $\nrlm$ and opposite in sign. On that respect, one may notice in figure \ref{sfig:ne:mH:Cs250} that the coupling $\nrle$ is instead positive for large scalar masses. In fact, since the $\delta a_e$ anomaly must be obtained from two loop contributions in the whole range of scalar masses\footnote{Since $\mA>\mH$, $\nH$ gives the dominant one loop contribution, which is positive: this precludes a one loop explanation of $\delta a_e^{\rm Exp,Cs}$. For $\delta a_e^{\rm Exp,Rb}$ a one loop explanation would require, in any case, values of $\nrle$ which clash with the perturbativity requirements on the Yukawa couplings and lepton flavour universality constraints.} there exists a linear relation between $\nrle$ and $\nrlm$ in this specific region: $\nrlm \simeq -13 \nrle$. This relation arises from the fact that the two loop contributions to $\delta a_{\ell}$ are linear in $\nrl{\ell}$, as previously mentioned; it can be seen in the lower part of figure \ref{sfig:nmu:ne:Cs250} inside the $1\sigma$ region. Departure from this straight line signals important one loop contributions to $\delta a_\mu$.

Similarly to the muon case, the dominant two loop contributions to $\delta a_e$ for large scalar masses are the top quark mediated ones. However, this dominance is not clear when considering smaller values of the scalar masses or, conversely, larger values of $\tb$, due to the $\tbinv$ suppression of the quark contributions. Indeed, the tau mediated contribution can be larger than the top mediated one. For this to occur, one might expect large values of both $|\nrlt|$ and $|\nrle|$, with $\nrlt$ and $\nrle$ having the same sign. Then one may have allowed regions with $\nrle<0$ which are absent for dominant top quark contributions.

We must stress that $\nrlt$ is rather unconstrained in this framework and, as we have just pointed out, there exist regions of the parameter space that require a sizable tau lepton coupling. These allowed regions might be helpful to explain the ATLAS excess in $pp \rightarrow S \rightarrow \tau^{+}\tau^{-}$, which certainly requires light neutral scalars that couple significantly to tau leptons.

\begin{figure*}[h]
\begin{center}
\subfloat[$\mH$ vs. $\tb$.\label{sfig:mH:tb:Cs250}]{\includegraphics[width=0.3\textwidth]{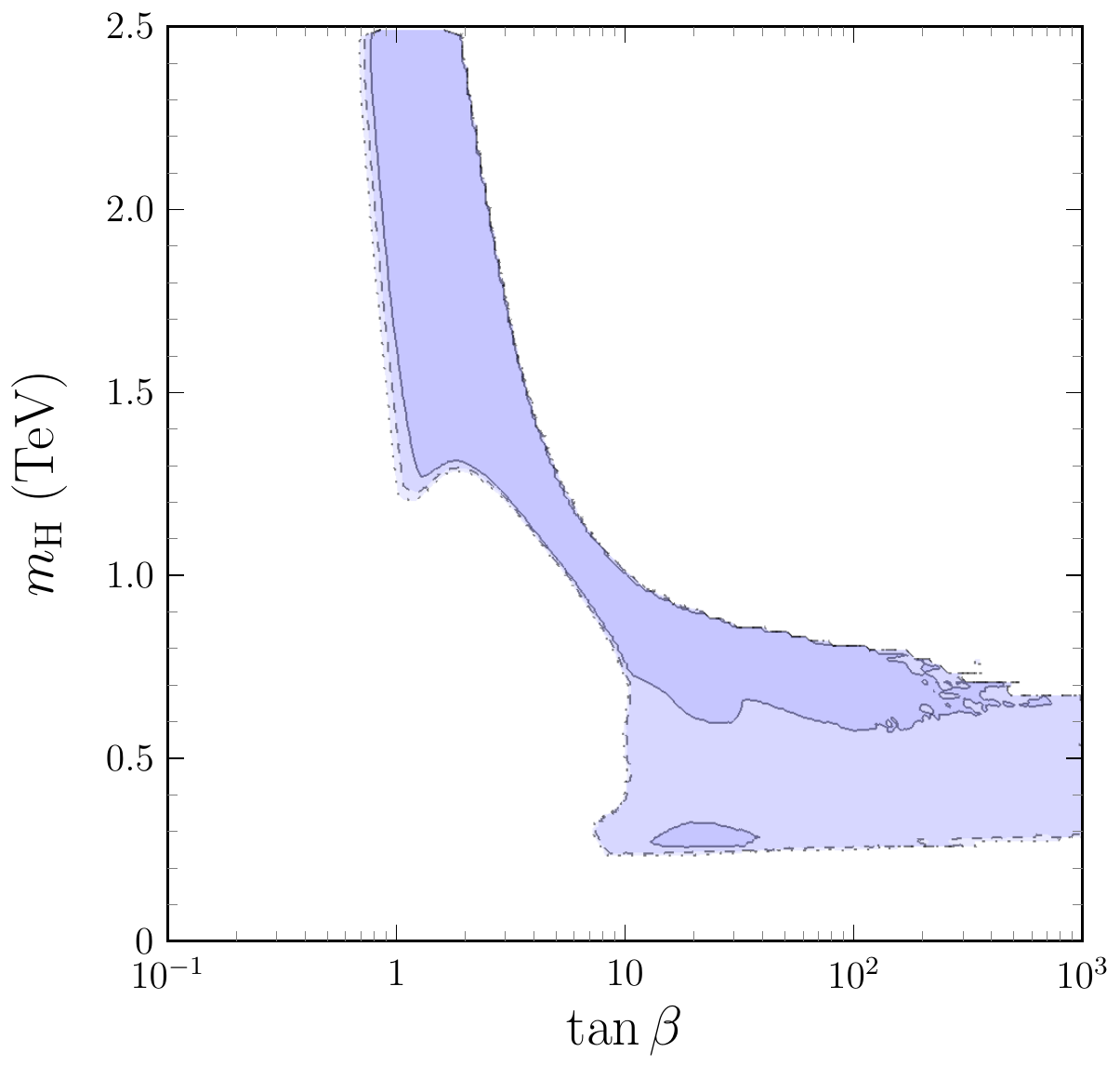}}\quad 
\subfloat[$\mA$ vs. $\mH$.\label{sfig:mA:mH:Cs250}]{\includegraphics[width=0.3\textwidth]{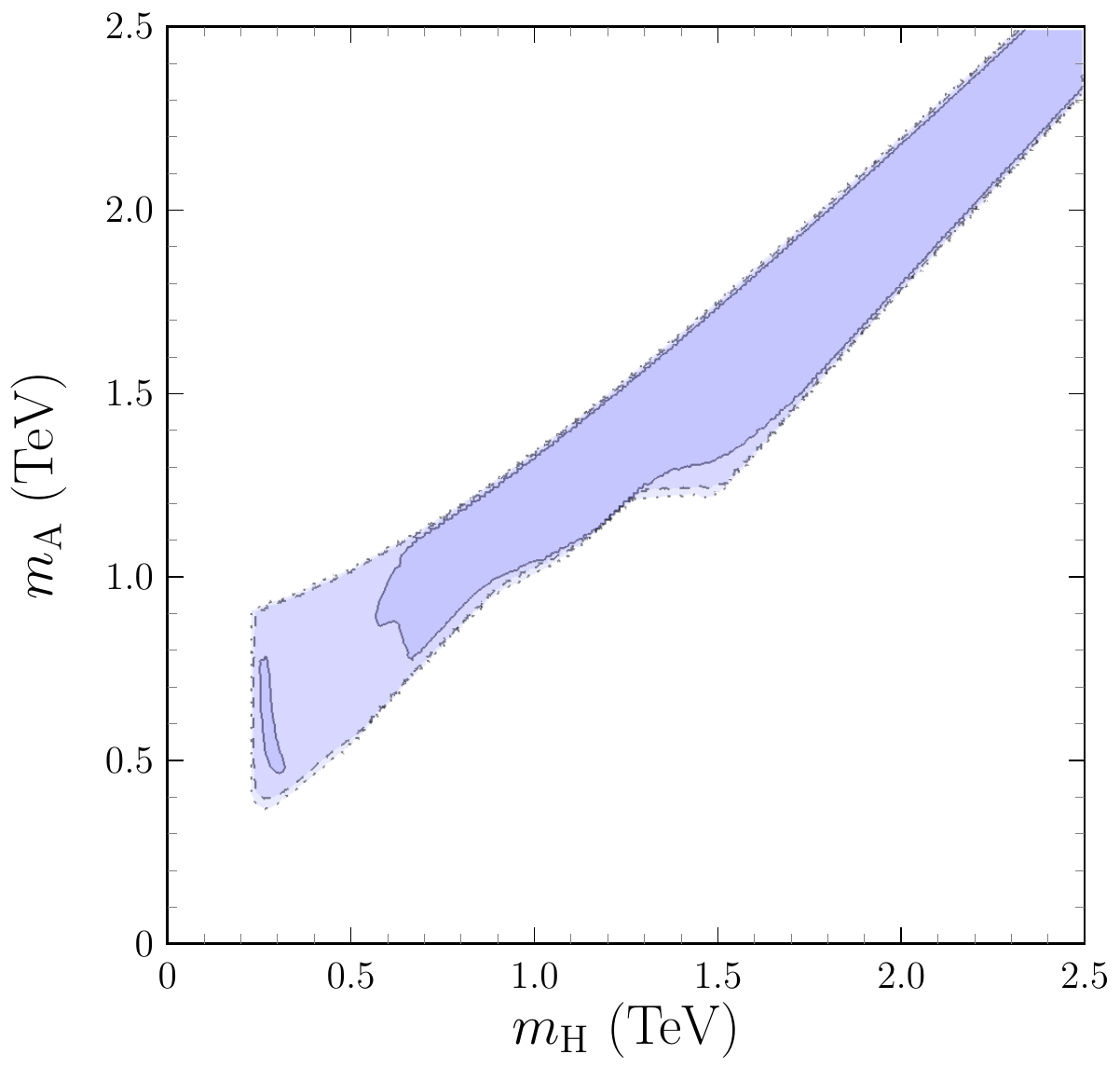}}\quad\subfloat[$\nrlm$ vs. $\mH$.\label{sfig:nmu:mH:Cs250}]{\includegraphics[width=0.3\textwidth]{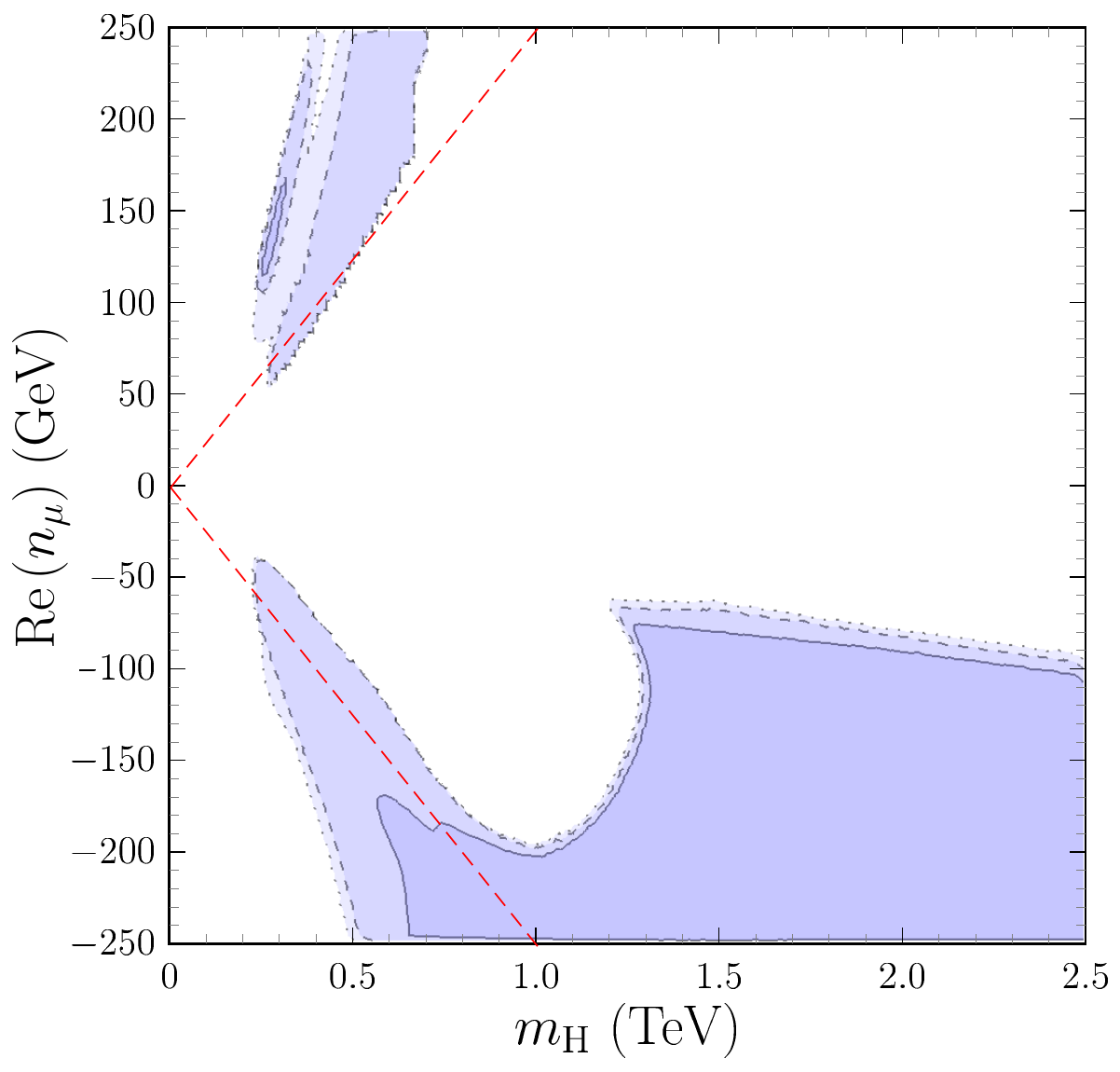}}
\caption{Allowed regions: relevant correlations involving $\mH$. The red dashed lines in figure \ref{sfig:nmu:mH:Cs250} correspond to $\nrlm=\pm\mH/4$.}\label{fig:01:Cs250}
\end{center}
\end{figure*}
\begin{figure*}[h!tb]
\begin{center}
\subfloat[$\nrle$ vs. $\mH$.\label{sfig:ne:mH:Cs250}]{\includegraphics[width=0.3\textwidth]{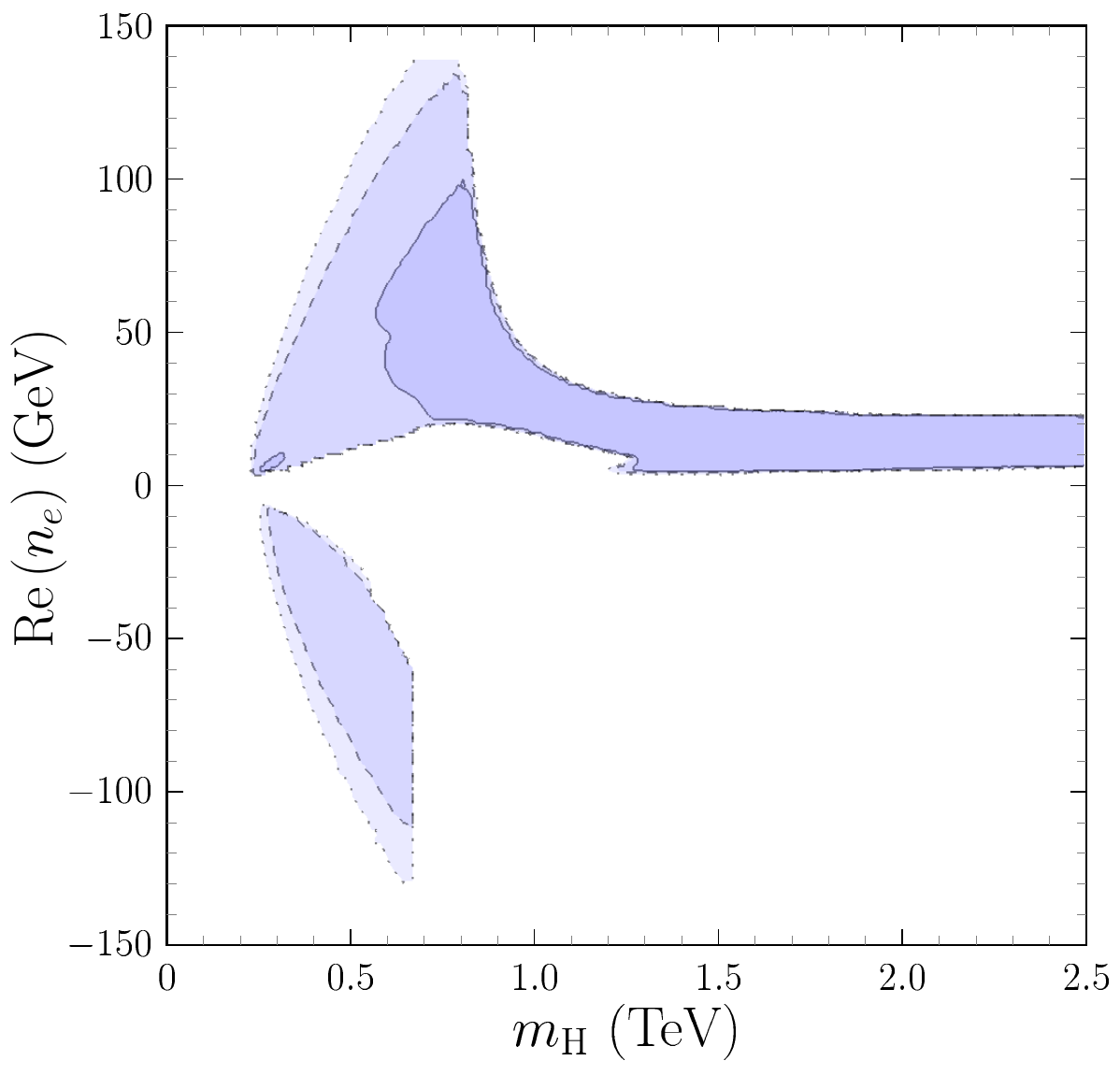}}\quad
\subfloat[$\nrlm$ vs. $\nrle$.\label{sfig:nmu:ne:Cs250}]{\includegraphics[width=0.3\textwidth]{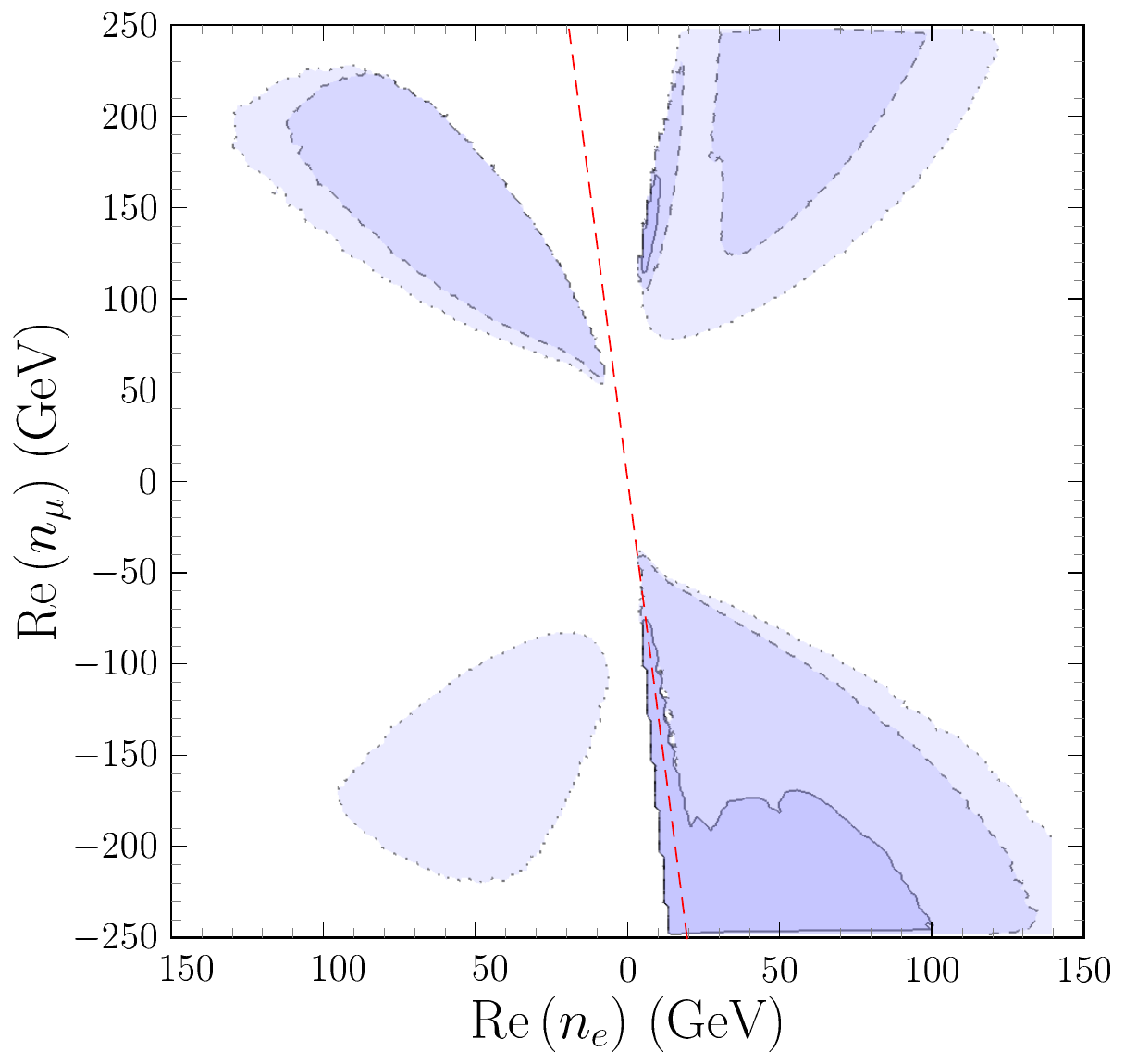}}\quad 
\subfloat[$\nrlt$ vs. $\nrle$.\label{sfig:ntau:ne:Cs250}]{\includegraphics[width=0.3\textwidth]{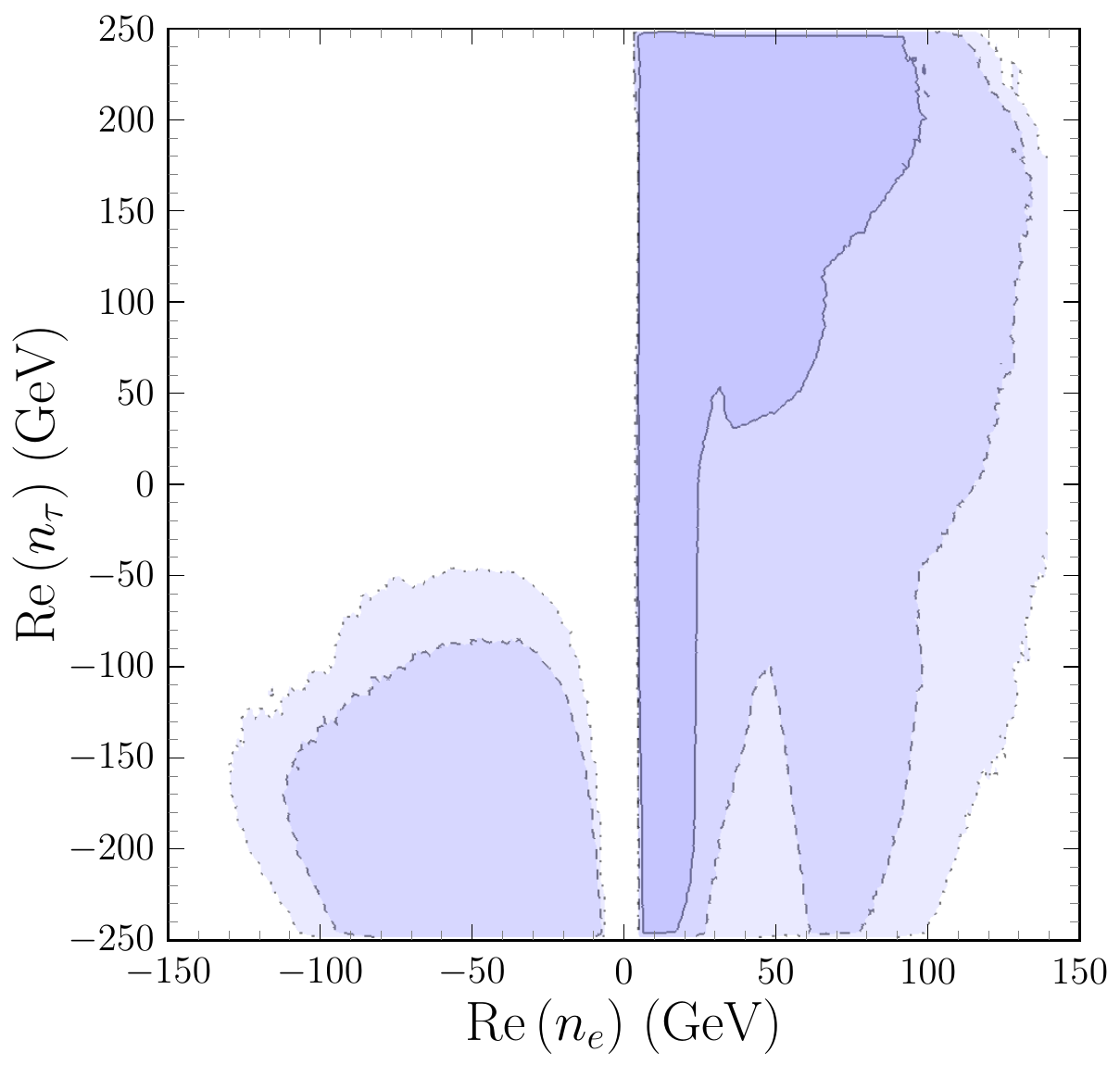}}
\caption{Allowed regions: relevant correlations involving $\nrle$. The red dashed line in figure \ref{sfig:nmu:ne:Cs250} corresponds to $\nrlm=-13\nrle$.}\label{fig:02:Cs250}
\end{center}
\end{figure*}
One might worry that large $\nl{\tau}$ values give large new contributions in processes such as $B\to D^{(\ast)}\tau\nu$, $B\to\tau\bar\tau$ or $B\to K^\ast \tau\bar\tau$. One can check that the $\tb$ factors arising from the quark couplings, together with the new scalar masses involved, effectively suppress these contributions despite large $\nl{\tau}$ values.

\section{The ATLAS excess in $pp\to S\to\tau^+\tau^-$\label{SEC:ATLAS}}
In \cite{ATLAS:2020zms} the ATLAS collaboration presented results concerning the search for heavy scalars decaying into two tau leptons over a mass range $[0.2;2.5]$ TeV. Although overall the data do not disagree significantly from the background prediction of the SM, the observed upper limits on the production cross section times branching fraction are larger than the corresponding expectation for masses in the $[0.2;0.6]$ TeV range. This happens in gluon-gluon fusion (ggF) production and in $b$-associated production. For the analyses presented in this work, we consider that an explanation of this excess is obtained when the production cross section times branching ratio $\sigma(pp\to S)\times \BR{S\to\tau^+\tau^-}$ for at least one $S$ in $\{\nH,\nA\}$ has a value between the expected ATLAS bound and the observed ATLAS bound for the corresponding mass of the scalar, that is 
\begin{equation}\label{eq:ATLAS:00}
[\text{Observed bound}]\geq \sigma(pp\to S)\times\BR{S\to\tau^+\tau^-}\geq [\text{Expected bound}].
\end{equation}
We may have (i) an explanation due to $\nH$ with $\mH\in[0.2;0.6]$ TeV without regard to $\mA$, (ii) an explanation due to $\nA$ with $\mA\in[0.2;0.6]$ TeV without regard to $\mH$. In addition, since (i) and (ii) are not mutually exclusive, we might even have ``a double explanation'', that is two local excesses for both $\mH,\mA\in[0.2;0.6]$ TeV.
One may worry that for $\mH\simeq \mA$ the approach is inconsistent on two respects. First, for a given production mode, $pp\to\nH\to\tau^+\tau^-$ and $pp\to\nA\to\tau^+\tau^-$ can in principle interfere, and then one should combine both processes at the amplitude level. As discussed in precedence, $\nH$ is a scalar and $\nA$ a pseudoscalar and thus there is no such interference. Second, although \refeq{eq:ATLAS:00} is fulfilled separately for both $S=\{\nH,\nA\}$, it may still happen that, for $\mH\simeq \mA$,
\begin{equation}
\sigma(pp\to \nH)\times\BR{\nH\to\tau^+\tau^-}+\sigma(pp\to \nA)\times\BR{\nA\to\tau^+\tau^-}\geq [\text{Observed bound}].
\end{equation}
In this sense, reproducing $\delta a_e$ and $\delta a_\mu$ (together with the constraints mentioned in section \ref{SEC:Paper1}) imposes $\mH\neq\mA$ for $\mH<800$ GeV, and this second source of concern is avoided. Our simple approach to an explanation of the ATLAS excess is thus safely consistent. 
Finally, the ATLAS results in \cite{ATLAS:2020zms} involve two different production modes: ggF and $b$-associated production. As discussed later, in subsection \ref{sSEC:NP:LHC}, it is important to clarify that the model can accommodate the deviation for gluon-gluon production but not for $b$-associated production and thus, in the following, we refer exclusively to $\sigma(pp\to S)_{\rm [ggF]}\times\BR{S\to\tau^+\tau^-}$ as ``the ATLAS excess''. Figure \ref{fig:ATLAS:00} shows this targeted excess region.
\begin{figure}[h!tb]
\begin{center}
\includegraphics[width=0.3\textwidth]{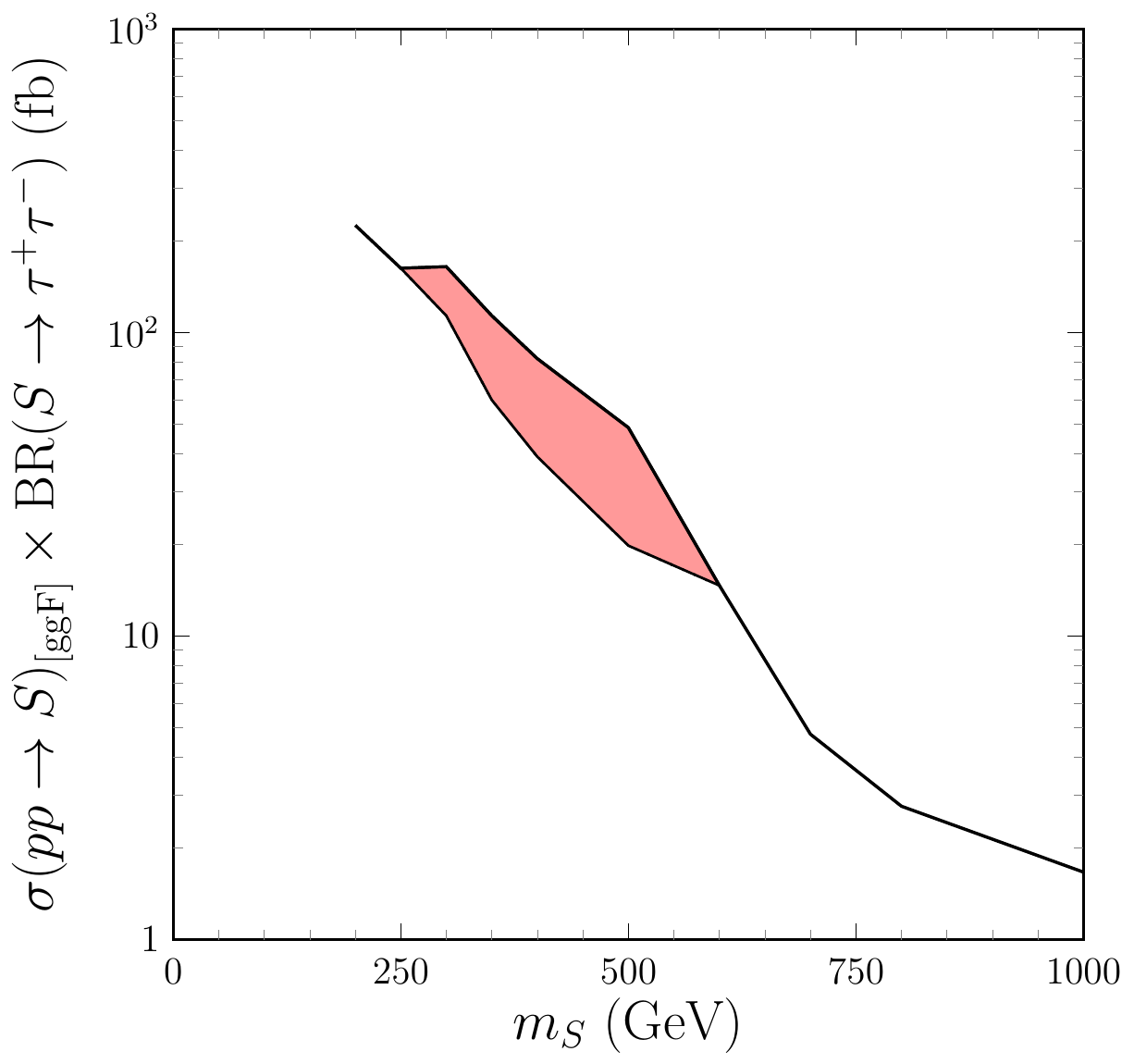} 
\caption{``ATLAS excess'': the colored region is limited above/below by the observed/expected bound on $\sigma(pp\to S)_{\rm [ggF]}\times\BR{S\to\tau^+\tau^-}$  vs. $m_S$. In our approach, the model is able to reproduce the excess when $\sigma(pp\to S)_{\rm [ggF]}\times\BR{S\to\tau^+\tau^-}$ vs. $m_S$, for $S=\nH$ and/or $\nA$, is inside the colored region, that is when there is, at least, a local excess with respect to the expected bound in the mass range of interest.\label{fig:ATLAS:00}}
\end{center}
\end{figure}

\section{Results\label{SEC:Results}}
Before addressing the results that we obtain through a detailed numerical analysis, a few comments are in order according to the
discussions in sections \ref{SEC:Paper1} and \ref{SEC:ATLAS} concerning previous results for this model and the ATLAS excess of interest.
\\ 
Besides the fact that $\mA>\mH$ for $\mA,\mH<1$ TeV, illustrated by figure \ref{sfig:mA:mH:Cs250}, it is to be noticed that masses $\mA,\mH<600$ GeV require $\tb> 8$, as figure \ref{sfig:mH:tb:Cs250} shows. In order to reproduce the ATLAS excess we certainly need $\sigma(pp\to S)_{\rm [ggF]}$ larger than some minimal value, but the leading ggF top mediated contribution is $\tb^{-2}$ suppressed: although large values of $\tb$ are required for $\delta a_e$, $\delta a_\mu$ when $\mH,\mA<600$ GeV, the ATLAS excess could in principle be easier to accommodate for small values of $\tb$. It also follows from the analysis in \cite{Botella:2022rte} that for $\tb> 8$ the appropriate $\delta a_\mu$ is obtained from a combination of one loop and two loop contributions, while $\delta a_e$ is two loop dominated. Within the two loop contributions, the ones with virtual quarks are $\tb^{-1}$ suppressed and there are regions in parameter space where large $|\nl{\tau}|$ can give contributions with virtual $\tau$'s of similar size, both in $\delta a_e$ and $\delta a_\mu$. This brings us to the second ingredient for the ATLAS excess: $\BR{S\to\tau^+\tau^-}$. The rate $\Gamma(S\to\tau^+\tau^-)$ is proportional to $|\nl{\tau}|^2$ and thus (i) a minimal $|\nl{\tau}|$ is necessary and (ii) large values of $|\nl{\tau}|$, bounded in any case by perturbativity requirements, can help accommodating the ATLAS excess.\\
All in all, from this short discussion we might expect that, with respect to the allowed regions emerging from the analyses in \cite{Botella:2022rte}, an explanation of the ATLAS excess will have a substantial direct impact on the allowed values of the new scalar masses, on the values of $\tb$ and $\nrlt$, and indirectly on the values of $\nrlm$ and $\nrle$. Furthermore, it might force a peculiar mixture of one loop and two loop contributions in $\delta a_\mu$ while the two loop contributions themselves, in both $\delta a_\mu$ and $\delta a_e$, can be a mixture of top quark and tau induced terms. Some interesting (additional) NP implications of these expectations are explored in section \ref{SEC:NP}.\\
The rest of this section is devoted to selected results of the detailed numerical analysis which incorporates all necessary ingredients. In figures \ref{fig:ATLAS:nl}-\ref{fig:ATLAS:obs} we follow the same darker to lighter coloring of the allowed regions in section \ref{SEC:Paper1}. Since the analysis in this section includes in addition the ATLAS excess, it is to be mentioned that the best fit here is only slightly worse than the best fit of the analysis in section \ref{SEC:Paper1}: it is within the 1-2$\sigma$ region of that analysis and thus the ATLAS excess is reproduced without spoiling the good agreement with all other observables. Let us start by showing some relevant correlations involving the lepton couplings $\nrl{\ell}$. From figure \ref{sfig:nrlt:nrle:AoPh2F}, where $\nrlt$ vs. $\nrle$ is shown, it is straightforward to check that large positive values of $\nrlt$, namely $\nrlt > 120$ GeV, are required to accommodate the ATLAS excess. On that respect, these large values of $\nrlt$ generate an important two loop $\tau$ induced contribution to the electron anomaly at the level of, or even larger than, the corresponding quark contribution. As previously mentioned, this would imply positive values of $\nrle$ as well, which is clearly illustrated in this figure.

Regarding the correlation $\nrlm$ vs. $\nrle$ presented in figure \ref{sfig:nrlm:nrle:AoPh2F}, it satisfies roughly the linear relation $\nrlm \simeq -7\nrle$, suggesting the presence of relevant two loop contributions in both $\delta a_\mu$ and $\delta a_e$. However, one may notice that the proportionality constant here is reduced by almost a factor 2 with respect to that appearing in the analogous linear relation of section \ref{SEC:Paper1}. In this sense, it is important to realize that, for light scalar masses, the muon anomaly also receives significant one loop contributions, so that two loop terms only account for approximately 50\% of the total $\delta a_\mu$. The latter might justify the previous reduction factor. All in all, as advanced before, it is clear that the explanation of the ATLAS excess has an indirect impact on the values of $\nrle$ and $\nrlm$ through the $(g-2)_\ell$ constraints.

Figure \ref{sfig:mH:tb:AoPh2F} shows $\mH$ vs. $\tb$. One can readily observe that the allowed range for $\mH$, namely $\mH \in [0.2;0.6]$ TeV, coincides precisely with the excess region and is related to $\tb \sim 10$ values, as required for a simultaneous explanation of both $\delta a_\mu^{\rm{Exp}}$ and $\delta a_e^{\rm{Exp,Cs}}$. On the other hand, figure \ref{sfig:mA:mH:AoPh2F} shows that $\mA$ lies in the range $\mA \in [0.4;1.0]$ TeV, with $\mA > \mH$, which obviously contains allowed values that do not correspond to the scalar masses giving rise to the excess. Those points definitely belong to the scenario where the ATLAS excess is explained through H contributions without regard to $\mA$. Since $\cH$ has no significant role in $(g-2)_{e,\mu}$ and in the ATLAS excess, we do not include results concerning $\mcH$; it is worth mentioning, however, that one has $\mcH\sim\mA$, mainly because of the oblique parameters constraint.

This behavior can be also observed in $\sigma(pp \to S)_{\rm{[ggF]}} \times \BR{S\to\tau^+\tau^-}$ vs. $m_S$, with $S = \{\nH, \nA\}$, presented in figures \ref{sfig:sigma:mH:AoPh2F} and \ref{sfig:sigma:mA:AoPh2F}. The allowed regions are mainly located above the expected bound and below the observed bound, and might correspond to an explanation of the excess in terms of H and/or A. However, there are still some allowed points below the expected bound both in the scalar and pseudoscalar contributions: those points certainly require an explanation of the excess due to A without regard to $\mH$ or due to H without regard to $\mA$, respectively.

\begin{figure}[h!tb]
\begin{center}
\subfloat[$\nrlt$ vs. $\nrle$.\label{sfig:nrlt:nrle:AoPh2F}]{\includegraphics[width=0.3\textwidth]{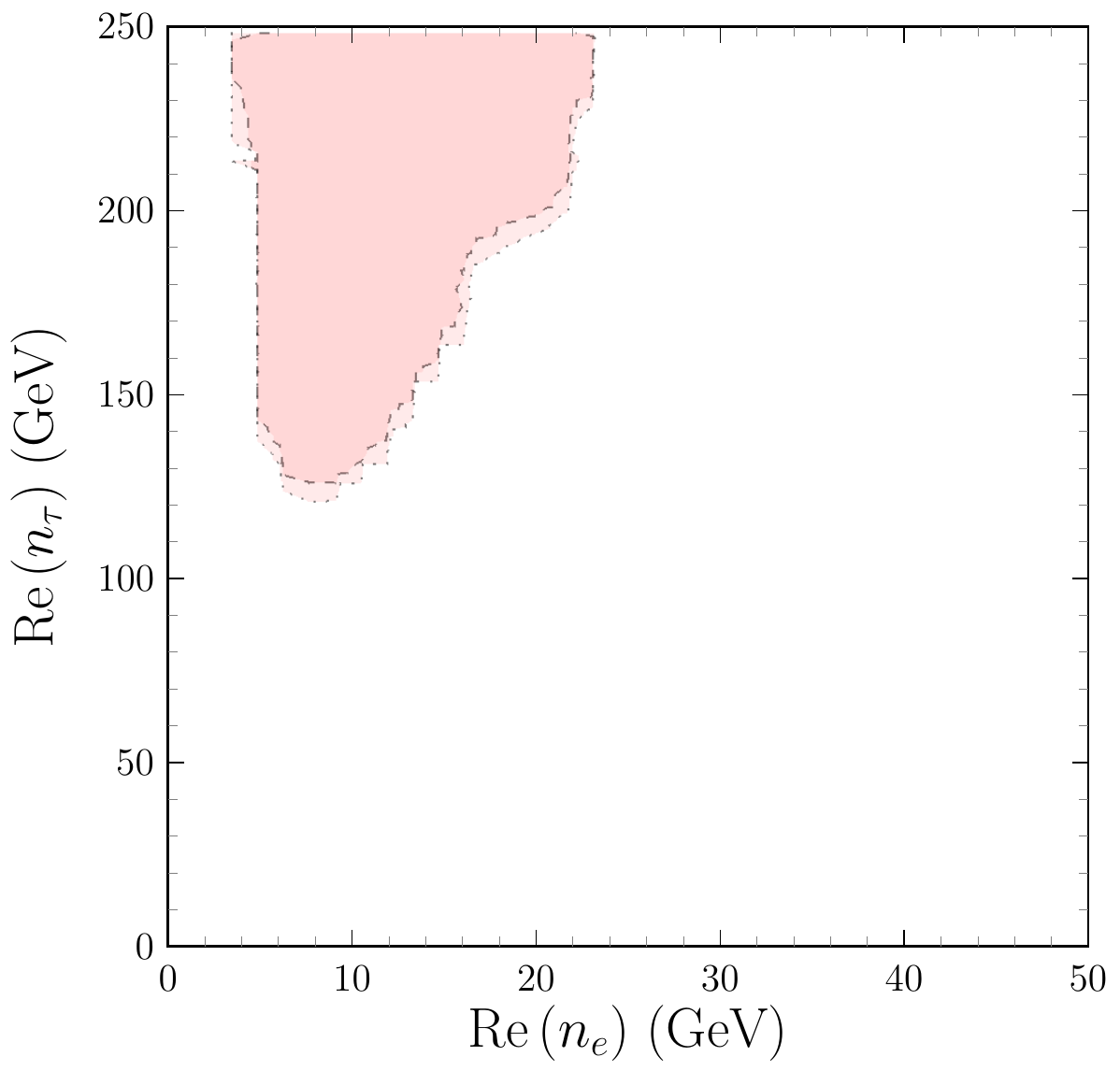}}\quad
\subfloat[$\nrlm$ vs. $\nrle$.\label{sfig:nrlm:nrle:AoPh2F}]{\includegraphics[width=0.3\textwidth]{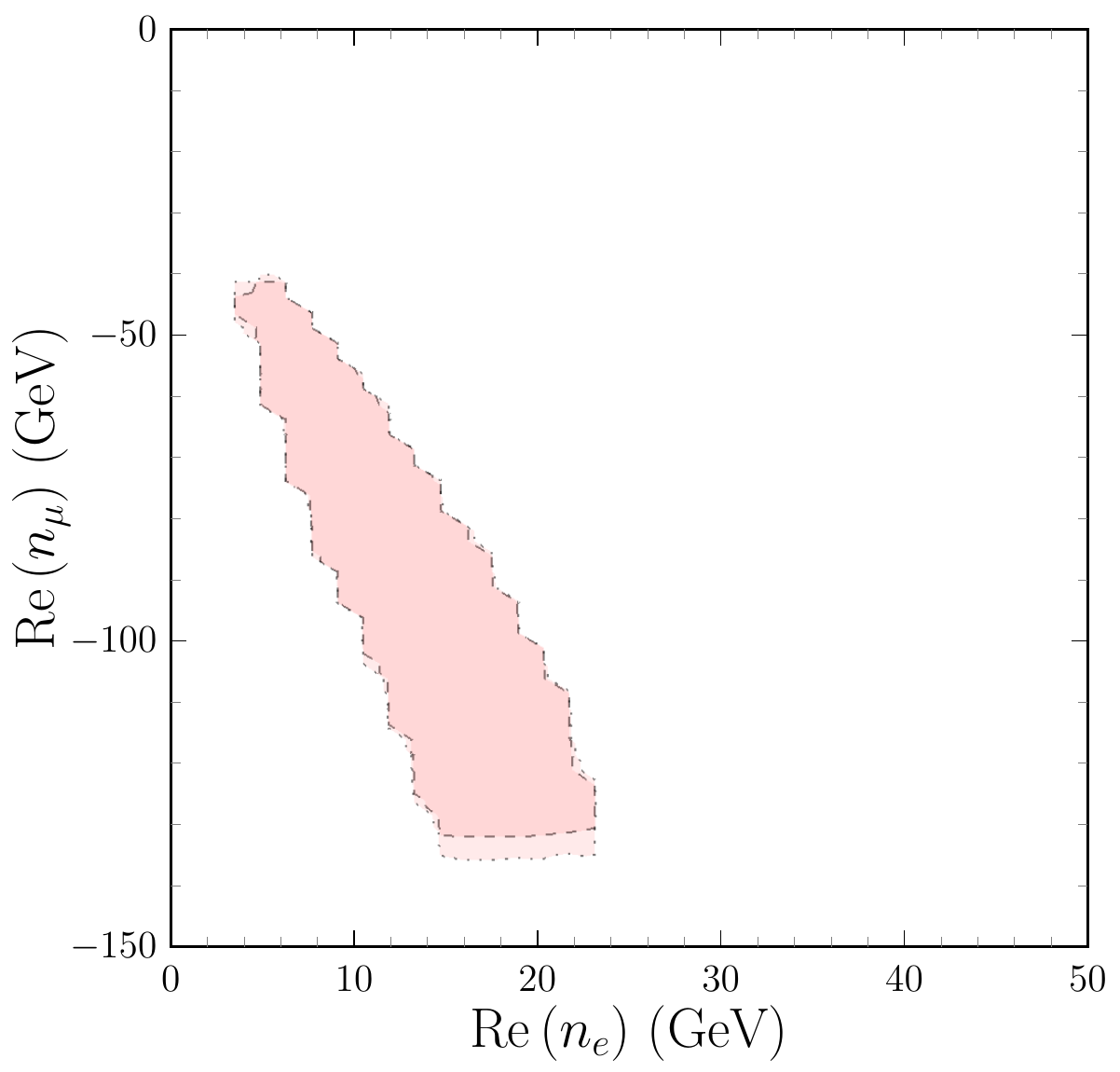}}
\caption{Allowed regions including the ATLAS excess: lepton couplings $\nrlt$, $\nrlm$ vs. $\nrle$.}\label{fig:ATLAS:nl}
\end{center}
\end{figure}
\begin{figure}[h!tb]
\begin{center}
\subfloat[$\mH$ vs. $\tb$.\label{sfig:mH:tb:AoPh2F}]{\includegraphics[width=0.3\textwidth]{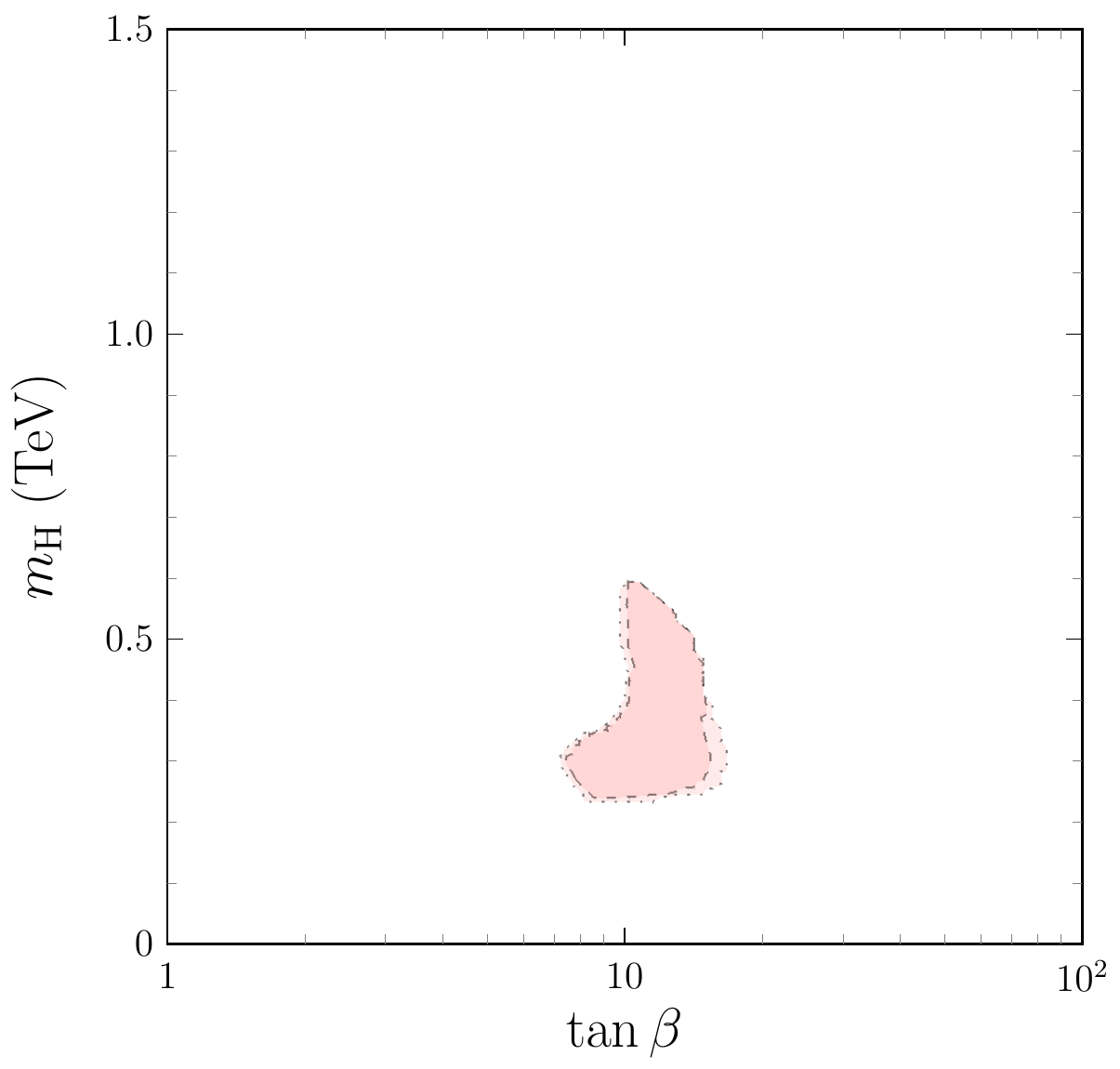}}\quad
\subfloat[$\mA$ vs. $\mH$.\label{sfig:mA:mH:AoPh2F}]{\includegraphics[width=0.3\textwidth]{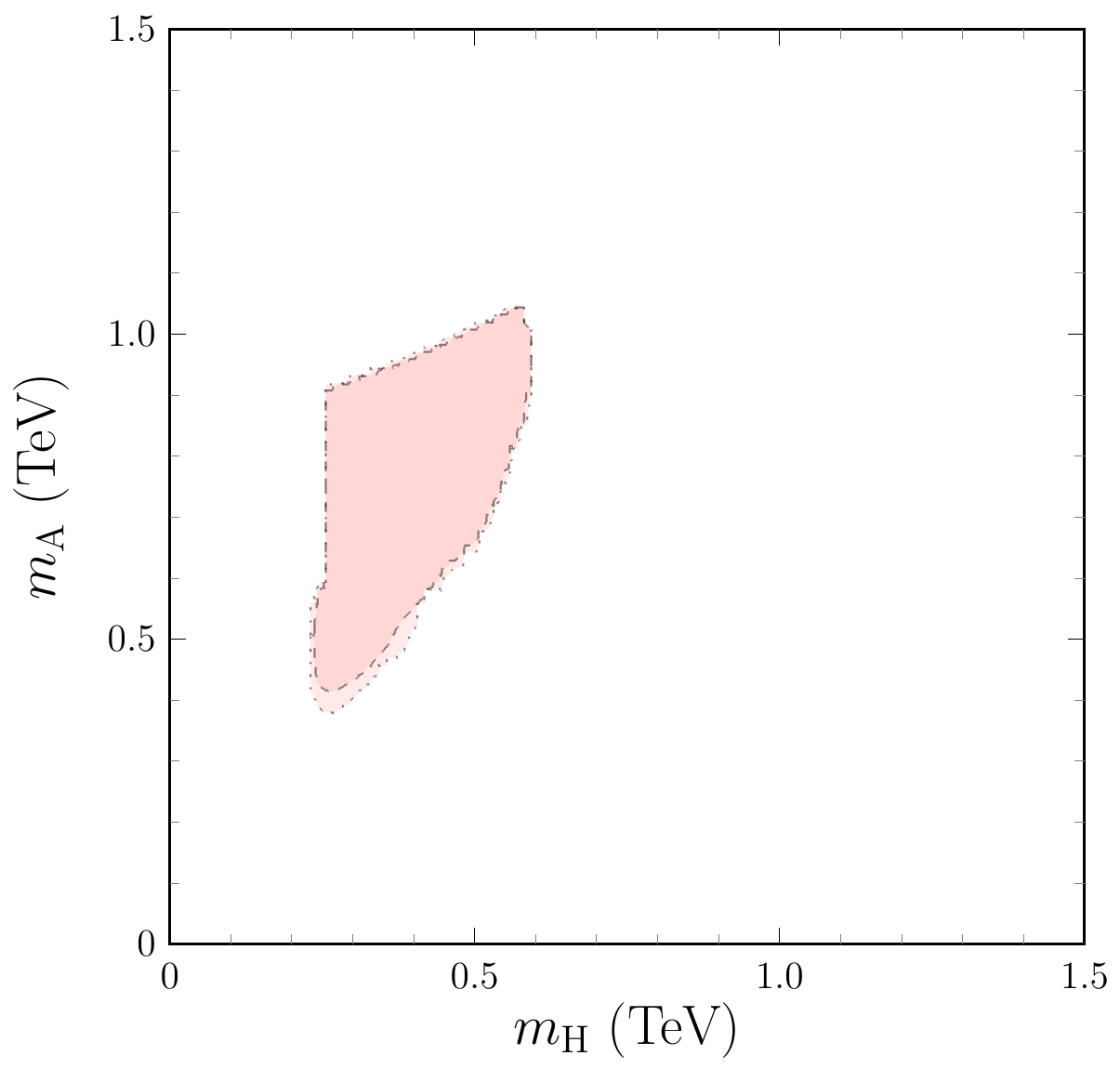}}
\caption{Allowed regions including the ATLAS excess: relevant correlations involving $\mH$.\label{fig:ATLAS:mH}}
\end{center}
\end{figure}
\begin{figure}[h!tb]
\begin{center}
\subfloat[For $S = \nH$.\label{sfig:sigma:mH:AoPh2F}]{\includegraphics[width=0.3\textwidth]{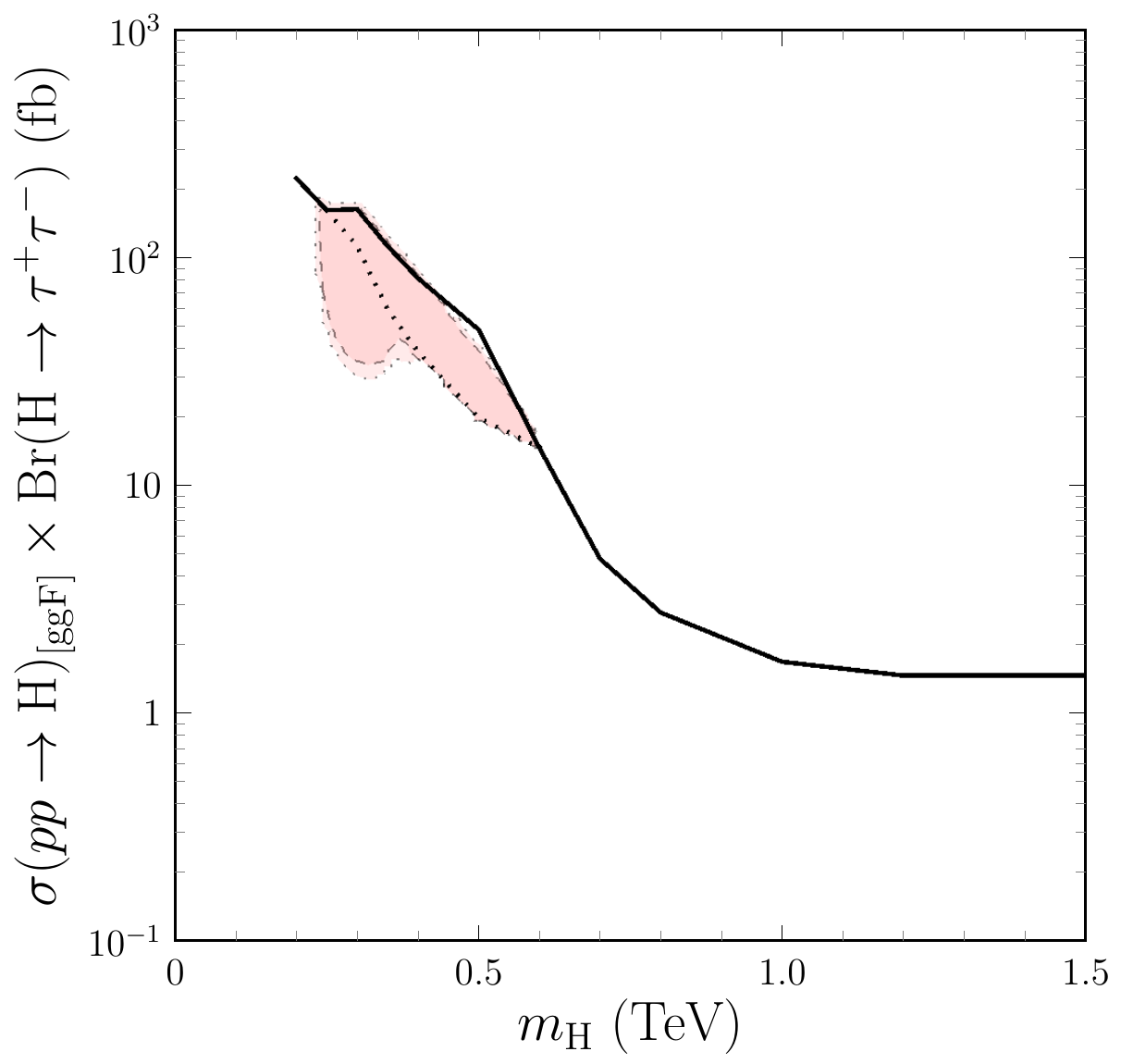}}\quad
\subfloat[For $S = \nA$.\label{sfig:sigma:mA:AoPh2F}]{\includegraphics[width=0.3\textwidth]{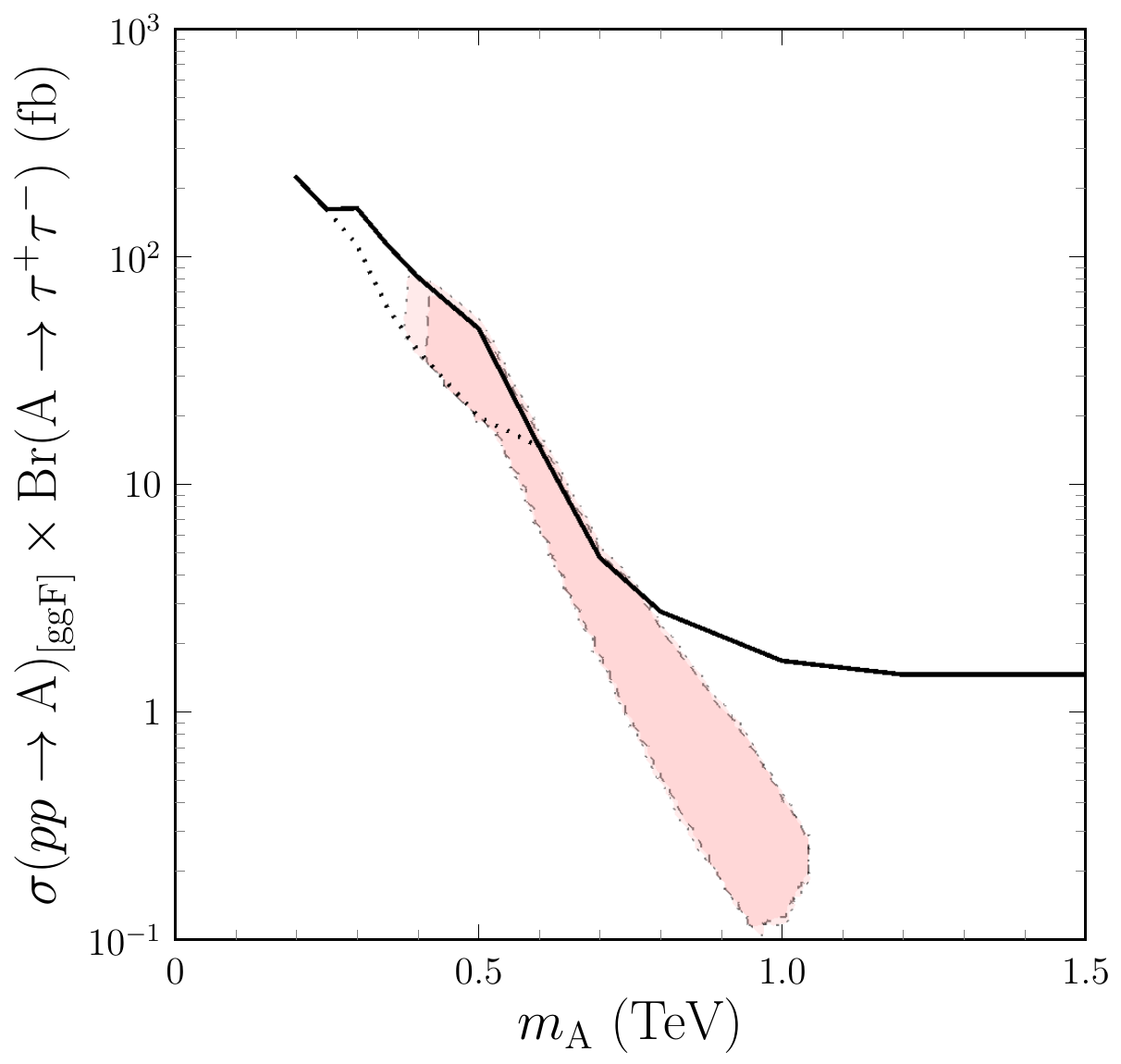}}
\caption{$\sigma(pp \rightarrow S)_{[\rm{ggF}]} \times \BR{S \rightarrow \tau^+\tau^-}$ vs. $\mS$ allowed regions including the ATLAS excess.\label{fig:ATLAS:obs}}
\end{center}
\end{figure}

\begin{table*}[h]
\begin{center}
 \begin{tabular}{|c|c|c|c|c|c|c|c|c|}\hline
  Point & $\mH$ & $\mA$ & $\mcH$ & $\tb$ & $\re{\mu_{12}^2}$ & $\nrle$ & $\nrlm$ & $\nrlt$\\ \hline\hline
  1 & $362$ & $775$ & $778$ & $10.7$ & $-1.19\times 10^4$ & $8.16$ & $-72.8$ & $222.5$\\ \hline
  2 & $277$ & $494$ & $502$ & $10.8$ & $-6.61\times 10^3$ & $9.12$ & $-58.0$ & $191.2$\\ \hline
  3 & $278$ & $539$ & $554$ & $7.93$ & $-8.90\times 10^5$ & $6.71$ & $-53.1$ & $189.7$\\ \hline
  \end{tabular}
 \caption{Example points: masses and $\nrl{\ell}$'s in GeV, $\re{\mu_{12}^2}$ in GeV$^2$.\label{tab:examples:pars}}
\end{center}
\end{table*}

\begin{table*}[h!]
\begin{center}
 \begin{tabular}{|c|c|c|c|c|c|c|c|c|c|c|}\cline{3-11}
 \multicolumn{2}{c}{}& \multicolumn{3}{|c}{1 loop} & \multicolumn{6}{|c|}{2 loop}\\ \hline
  Point & $\delta a_e \times 10^{13}$ & $\nH$ & $\nA$ & $\cH$ & $t\nH$ & $t\nA$ & $\tau\nH$ & $\tau\nA$ & $\mu\nH$ & $\mu\nA$\\ \hline\hline
  1 & $-7.24$ & $0$ & $0$ & $0$ & $0.307$ & $0.199$ & $0.779$ & $-0.259$ & $-0.036$ & $0.010$\\ \hline
  2 & $-9.84$ & $-0.001$ & $0$ & $0$ & $0.310$ & $0.258$ & $0.845$ & $-0.388$ & $-0.038$ & $0.015$\\ \hline
  3 & $-8.87$ & $0$ & $0$ & $0$ & $0.347$ & $0.264$ & $0.683$ & $-0.271$ & $-0.028$ & $0.010$\\ \hline
 \end{tabular}
 \caption{Example points: $\delta a_e \times 10^{13}$ values. Columns 3 to 11 show the relative contributions of the different one and two loop terms to the value of $\delta a_e \times 10^{13}$ in the second column.\label{tab:examples:dae}}
\end{center}
\end{table*}

\begin{table*}[h!]
\begin{center}
 \begin{tabular}{|c|c|c|c|c|c|c|c|c|c|c|}\cline{3-11}
 \multicolumn{2}{c}{}& \multicolumn{3}{|c}{1 loop} & \multicolumn{6}{|c|}{2 loop}\\ \hline
  Point & $\delta a_\mu \times 10^9$ & $\nH$ & $\nA$ & $\cH$ & $t\nH$ & $t\nA$ & $\tau\nH$ & $\tau\nA$ & $\mu\nH$ & $\mu\nA$\\ \hline\hline
  1 & $2.42$ & $0.588$ & $-0.136$ & $-0.003$ & $0.169$ & $0.110$ & $0.429$ & $-0.143$ & $-0.020$ & $0.006$\\ \hline
  2 & $2.29$ & $0.650$ & $-0.218$ & $-0.005$ & $0.175$ & $0.146$ & $0.477$ & $-0.219$ & $-0.021$ & $0.009$\\ \hline
  3 & $2.35$ & $0.531$ & $-0.147$ & $-0.003$ & $0.214$ & $0.163$ & $0.422$ & $-0.168$ & $-0.017$ & $0.006$\\ \hline
 \end{tabular}
 \caption{Example points: $\delta a_\mu \times 10^9$ values. Columns 3 to 11 show the relative contributions of the different one and two loop terms to the value of $\delta a_\mu \times 10^9$ in the second column.\label{tab:examples:damu}}
\end{center}
\end{table*}

\begin{table*}[h!]
\begin{center}
 \begin{tabular}{|c|c|c|c|c|c|c|}\hline
  Point & $S\to$ & $e\bar e$ & $\mu\bar\mu$ & $\tau\bar\tau$ & $t\bar t$ & $\nH Z$ \\ \hline\hline
 \multirow{2}{*}{1} & $\nH$ & $1.2\times 10^{-3}$ & $0.096$ & $0.902$ & $7\times 10^{-4}$ & $-$\\ \cline{2-7}
 & $\nA$ & $0.0004$ & $0.028$ & $0.265$ & $0.004$ & $0.703$ \\ \hline
 \multirow{2}{*}{2} & $\nH$ & $0.002$ & $0.084$ & $0.914$ & $-$ & $-$\\ \cline{2-7}
 & $\nA$ & $0.001$ & $0.049$ & $0.528$ & $0.008$ & $0.415$ \\ \hline
 \multirow{2}{*}{3} & $\nH$ & $0.001$ & $0.073$ & $0.926$ & $-$ & $-$ \\ \cline{2-7}
 & $\nA$ & $0.0005$ & $0.033$ & $0.415$ & $0.012$ & $0.540$ \\ \hline
 \end{tabular}
 \caption{Example points: decay branching ratios of $\nH$ and $\nA$.\label{tab:examples:BR}}
\end{center}
\end{table*}

\begin{table}[h!]
\begin{center}
 \begin{tabular}{|c|c|c|c|c|c|}\cline{3-5}
\multicolumn{2}{c}{} & \multicolumn{3}{|c|}{$\sigma(pp\to S)_{\rm [ggF]}\times \BR{S\to\tau^+\tau^-}$ (fb)} & \multicolumn{1}{c}{}\\ \hline
\multicolumn{2}{|c|}{Point}  & Value & Expected bound & Observed bound & Excess\\ \hline\hline
 \multirow{2}{*}{1} & $\nH$ & $83.6$ & $54.9$ & $106.0$ & \checkmark\\ \cline{2-6}
 & $\nA$ & $1.50$ & $3.2$ & $3.2$ & --\\ \hline
 \multirow{2}{*}{2} & $\nH$ & $84.0$ & $136.0$ & $164.0$ & --\\ \cline{2-6}
 & $\nA$ & $27.3$ & $20.8$ & $50.4$ & \checkmark\\ \hline
 \multirow{2}{*}{3} & $\nH$ & $160.1$ & $135.4$ & $164.0$ & \checkmark\\ \cline{2-6}
 & $\nA$ & $27.6$ & $17.6$ & $31.1$ & \checkmark\\ \hline
 \end{tabular}
 \caption{Example points: values of $\sigma(pp\to S)_{\rm [ggF]}\times \BR{S\to\tau^+\tau^-}$ are shown for both $\nH$ and $\nA$, together with the expected and observed bounds for the corresponding value of the mass.\label{tab:examples:tautau}}
\end{center}
\end{table}
Tables \ref{tab:examples:pars}--\ref{tab:examples:tautau} show 3 detailed example points to further illustrate the results of the analysis. Besides the parameters in table \ref{tab:examples:pars}, the different contributions to $\delta a_e$ and $\delta a_\mu$ are shown in detail in tables \ref{tab:examples:dae} and \ref{tab:examples:damu}. One can readily observe that $\delta a_e$ is obtained with two loop contributions, with similar size for virtual top quarks and tau leptons $t\nH+t\nA\sim \tau\nH+\tau\nA$ (including the partial $\tau\nH$ vs. $\tau\nA$ cancellation). One can also observe similar two loop top and tau contributions in $\delta a_\mu$, with the important difference that, contrary to $\delta a_e$, these two loop contributions account for 50-60\% of $\delta a_\mu$ with the remaining 50-40\% coming from one loop contributions (including the partial $\nH$ vs. $\nA$ cancellation). Table \ref{tab:examples:BR} shows the most relevant decay channels of the neutral scalars $\nH$ and $\nA$. The $\nrl{\ell}$ couplings in table \ref{tab:examples:pars} clearly shape the strong hierarchy in the decay branching ratios of $\nH$ which are dominated by $\nH\to\tau^+\tau^-$. For $\nA$, there is also a strong hierarchy among the different $\BR{\nA\to\ell^+\ell^-}$, but there is in addition a large, even dominating, decay $\nA\to\nH Z$. In table \ref{tab:examples:tautau} the expected and observed bounds on $\sigma(pp\to S)_{\rm [ggF]}\times \BR{S\to\tau^+\tau^-}$ are shown together with its corresponding value for the example points: these examples are chosen to illustrate that one may have an excess due to $\nH$ but not $\nA$ (point 1), due to $\nA$ but not $\nH$ (point 2), and due to both $\nH$ and $\nA$ (point 3).

We close this section with a comment on $\delta a_e^{\rm Exp}$. As discussed in section \ref{SEC:Intro}, one can consider the two different determinations of $a_e$ in \refeq{eq:electron:anomaly}. A detailed analysis of the effect of the value of $\delta a_e^\mathrm{Exp}$ can be found in \cite{Botella:2022rte}, where no $pp\to S\to \tau^+\tau^-$ excess was considered. In this work we have only shown results for $\delta a_e^{\rm Exp,Cs}$. Since the analysis shows that the allowed regions are rather constrained, it is important to consider what is the effect of changing the $\delta a_e^{\rm Exp}$ assumption from $\delta a_e^{\rm Exp,Cs}$ to $\delta a_e^{\rm Exp,Rb}$. For each allowed point, the simple change $\nrle\mapsto \frac{\delta a_e^{\rm Exp,Cs}}{\delta a_e^{\rm Exp,Rb}}\nrle$ with no other parameter change is able to accommodate the new $\delta a_e^{\rm Exp}$ value owing to the form of the dominant two loop contribution. The non-trivial aspect of this change is that it does not produce significant modifications in other constraints (in agreement with \cite{Botella:2022rte}) and thus all relevant aspects of the analysis and discussion would be essentially unaltered.

\section{Further New Physics hints\label{SEC:NP}}
%
In this section we discuss additional NP aspects. In subsection \ref{sSEC:NP:LHC} we analyse LHC hints other than the excess in $\sigma(pp\to S)_{\rm [ggF]}\times\BR{S\to\tau^+\tau^-}$ which may be of interest, but which cannot be simultaneously accommodated within this model. In subsection \ref{sSEC:NP:Implications} we discuss the consequences of accommodating the excess in $\sigma(pp\to S)_{\rm [ggF]}\times\BR{S\to\tau^+\tau^-}$ in other observables of particular interest to uncover the presence of NP.

\subsection{Other LHC excesses\label{sSEC:NP:LHC}}
%
Besides the ATLAS excess in $\sigma(pp\to S)_{\rm [ggF]}\times\BR{S\to\tau^+\tau^-}$, other potential hints of NP have been considered recently \cite{Richard:2020cav,Arganda:2021yms,Biekotter:2021qbc,Connell:2023jqq}. We discuss in the following why the ATLAS excess in $\sigma(pp\to S)_{[b-{\rm assoc}]}\times\BR{S\to\tau^+\tau^-}$ \cite{ATLAS:2020zms} and the CMS excess in ditop production \cite{CMS:2019pzc} are not compatible with the ATLAS excess in $\sigma(pp\to S)_{\rm [ggF]}\times\BR{S\to\tau^+\tau^-}$ within the model considered here.
\begin{itemize}
 \item For $\sigma(pp\to S)_{[b-{\rm assoc}]}\times\BR{S\to\tau^+\tau^-}$ it is important to notice that, since the quark Yukawa couplings correspond to a type I 2HDM, the $\tb$ dependence of top and bottom quark couplings to $\nH$ and $\nA$ is identical (see \refeq{eq:neutral:couplings}). It follows that $\sigma(pp\to S)_{\rm [ggF]}=\tb^{-2}\sigma(pp\to S)_{\rm [ggF]}^{\rm SM}$ and $\sigma(pp\to S)_{[b-{\rm assoc}]}=\tb^{-2}\sigma(pp\to S)_{[b-{\rm assoc}]}^{\rm SM}$, where $\sigma(pp\to S)_{\rm [ggF]}^{\rm SM}$ and $\sigma(pp\to S)_{[b-{\rm assoc}]}^{\rm SM}$ are the cross sections corresponding to a Higgs-like scalar (or pseudoscalar) with couplings to fermions $\sim \frac{m_f}{\vev{}}$. Then,
 \begin{equation}\label{eq:ggF:b:ATLAS}
  \frac{\sigma(pp\to S)_{\rm [ggF]}\times\BR{S\to\tau^+\tau^-}}{\sigma(pp\to S)_{[b-{\rm assoc}]}\times\BR{S\to\tau^+\tau^-}}=\frac{\sigma(pp\to S)_{\rm [ggF]}^{\rm SM}}{\sigma(pp\to S)_{[b-{\rm assoc}]}^{\rm SM}}\,.
 \end{equation}
From detailed calculations \cite{LHCHiggsCrossSectionWorkingGroup:2016ypw}, one can check that for $m_S\in[0.2;0.6]$ TeV, the right hand side has values in the $10^{3}-10^{4}$ range while for a simultaneous explanation of both ATLAS excesses the left hand side should take values of order 1: the excess for $b$-associated production cannot be accommodated once the ggF production excess is reproduced. One can observe that the impossibility to explain simultaneously the excesses with both production modes can be circumvented in a model with a type II 2HDM quark sector Yukawa couplings: instead of a $\tb$ independent right hand side in \refeq{eq:ggF:b:ATLAS}, one would have a $\tb^4$ factor in the denominator which, for $\tb\sim 10$, is in the right ballpark to match the left hand side. However, as discussed in \cite{Botella:2020xzf}, this possibility is barred by universality constraints in decays of pseudoscalar mesons.
\item For the CMS ditop excess in \cite{CMS:2019pzc}, the excess arises from the SM t-channel $gg\to t\bar t$ production together with ggF $gg\to S\to t\bar t$. Although the SM-NP interference may be helpful to enhance deviations from SM expectations, in the present scenario the NP process $gg\to S\to t\bar t$ has a $\tb^{-2}$ suppression at the amplitude level. Since $\tb>8$ is required for scalar masses $\mH,\mA\in[0.2;0.6]$ TeV, it is clear that the CMS ditop excess cannot be simultaneously accommodated.\footnote{For example, in the model independent analyses shown in \cite{CMS:2019pzc}, top-scalar couplings in the $\frac{m_t}{\vev{}}$ ballpark are required; in the model considered here, these couplings are $\frac{m_t}{\vev{}}\tbinv$ with $\tb>8$.}
\end{itemize}

\subsection{New Physics implications\label{sSEC:NP:Implications}}

\underline{\bf $\tau$ anomalous magnetic moment}\\

Besides the previous observables, another NP avenue of interest is the anomalous magnetic moment of the $\tau$ lepton. Although the current PDG bound \cite{ParticleDataGroup:2020ssz} is not yet at the sensitivity level of the Schwinger term $\frac{\alpha}{2\pi}\sim 10^{-3}$, the situation is improving \cite{Beresford:2019gww,Dyndal:2020yen,Jofrehei:2022bwh}. In the present scenario, considering the results of section \ref{SEC:Results}, in particular $\tb\sim 10$ and $\mcH,\mA>\mH$, one can expect that the NP contribution to $a_\tau$ is dominated by the one loop $\nH$ mediated term (an expectation that the complete numerical analysis fully confirms). For $\mH\sim 300$ GeV and $\nrlt\sim 200$ GeV, we have
\begin{equation}\label{eq:dat:1loop:H}
 \delta a_\tau\simeq \frac{1}{8\pi^2}\left(\frac{m_\tau \nrlt}{\vev{}\mH}\right)^2\left(-\frac{7}{6}-2\ln\left(\frac{m_\tau}{\mH}\right)\right)\simeq 2.7\times 10^{-6}\,.
\end{equation}
Fortunately, as discussed in detail in \cite{Crivellin:2021spu} (see also \cite{Bernabeu:2007rr}), sensitivities at the $10^{-6}$ level can be achieved at Belle-II through cross sections and polarization observables. With these considerations in mind, figure \ref{FIG:dat} shows results of the analysis in section \ref{SEC:Results} concerning $\delta a_\tau$.

\begin{figure}[!htb]
\begin{center}
\subfloat[$\delta a_\tau$ vs. $\nrlt$.\label{sFIG:dat:nt}]{\includegraphics[width=0.3\textwidth]{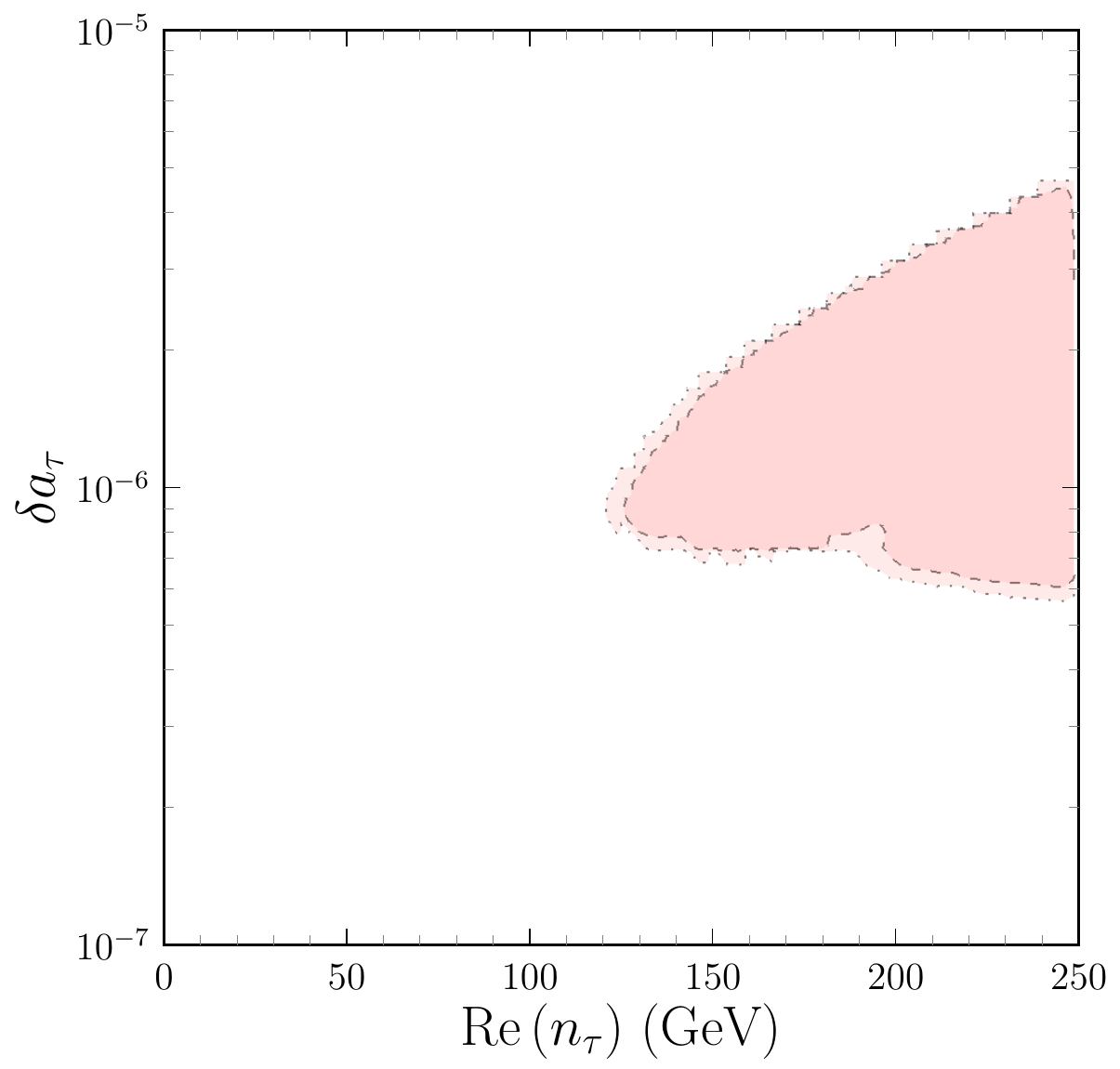}}\qquad
\subfloat[$\delta a_\tau$ vs. $\mH$.\label{sFIG:dat:MH}]{\includegraphics[width=0.3\textwidth]{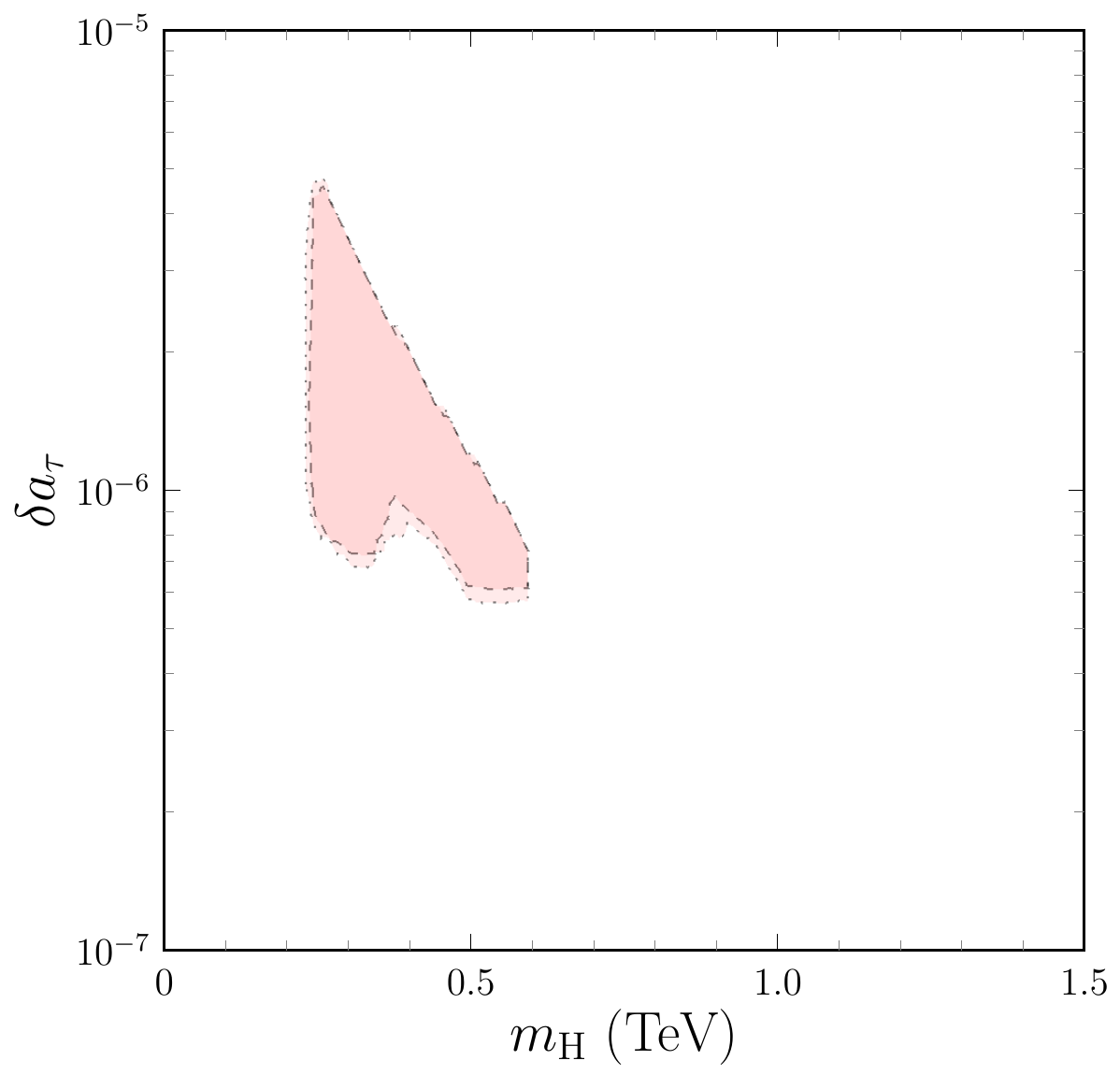}}
\caption{$\delta a_\tau$ allowed regions.\label{FIG:dat}}
\end{center}
\end{figure}
One can draw two important lessons: (i) $\delta a_\tau$ must necessarily have a value in the $[6\times 10^{-7};4\times 10^{-6}]$ range; (ii) although the largest values $\delta a_\tau\simeq 4\times 10^{-6}$ can only be obtained for $\nrlt\simeq\mH\simeq \vev{}$, we have values $\delta a_\tau\sim 10^{-6}$ over the entire allowed range $\vev{}/2<\nrlt<\vev{}$.\\ 

\noindent \underline{\bf CDF measurement of the mass of the $W$ boson}\\

Concerning the CDF measurement of $M_W$ in reference \cite{CDF:2022hxs}, the deviation in $M_W$ from the SM expectation obtained from analyses of electroweak precision observables can be translated into changes in the oblique parameters $S$, $T$ and $U$ \cite{deBlas:2022hdk,Lu:2022bgw}. One can thus explore the possibility to accommodate the CDF measurement by changing the oblique parameter values in the corresponding constraint used in the analysis. We consider two different scenarios, as done in \cite{Botella:2022rte}: 
\begin{itemize}
\item the ``conservative scenario'' in \cite{deBlas:2022hdk} which combines previous measurements with the CDF one, giving
 \begin{equation}\label{eq:DSDT:MW:cons}
  \Delta S=0.086\pm 0.077,\quad \Delta T=0.177\pm 0.070,\quad \rho=0.89\,,
 \end{equation}
 \item the scenario in \cite{Lu:2022bgw} which solely uses the CDF value and gives
 \begin{equation}\label{eq:DSDT:MW:chin}
  \Delta S=0.15\pm 0.08,\quad \Delta T=0.27\pm 0.06,\quad \rho=0.93\,.
 \end{equation}
\end{itemize}
$\Delta S\equiv S-S_{\rm SM}$ and $\Delta T\equiv T-T_{\rm SM}$ are the differences with respect to SM values.
Figures \ref{sFIG:DSDT:MW:cons} and \ref{sFIG:DSDT:MW:chin} show $\Delta S$ vs. $\Delta T$ allowed regions corresponding to \refeqs{eq:DSDT:MW:cons} and \eqref{eq:DSDT:MW:chin}, respectively. One can observe, as in \cite{Botella:2022rte}, that the model is able to accommodate better the $(\Delta S,\Delta T)\neq (0,0)$ values of the ``conservative scenario''. As figure \ref{FIG:DSDT:MW} clearly shows, this is due to the fact that in this model one can have $\Delta T$ at the $0.1$ to $0.2$ level, but nothing comparable in $\Delta S$. As was already noticed in \cite{Botella:2022rte}, the modified oblique parameter constraint does not have a significant impact on the allowed regions of parameters such as $\tb$ or the different $\nl{\ell}$'s (in fact, $\Delta S$ and $\Delta T$ do not depend directly on these parameters).
\begin{figure}
\begin{center}
\subfloat[$(\Delta T,\Delta S)$ in \refeq{eq:DSDT:MW:cons}.\label{sFIG:DSDT:MW:cons}]{\includegraphics[width=0.3\textwidth]{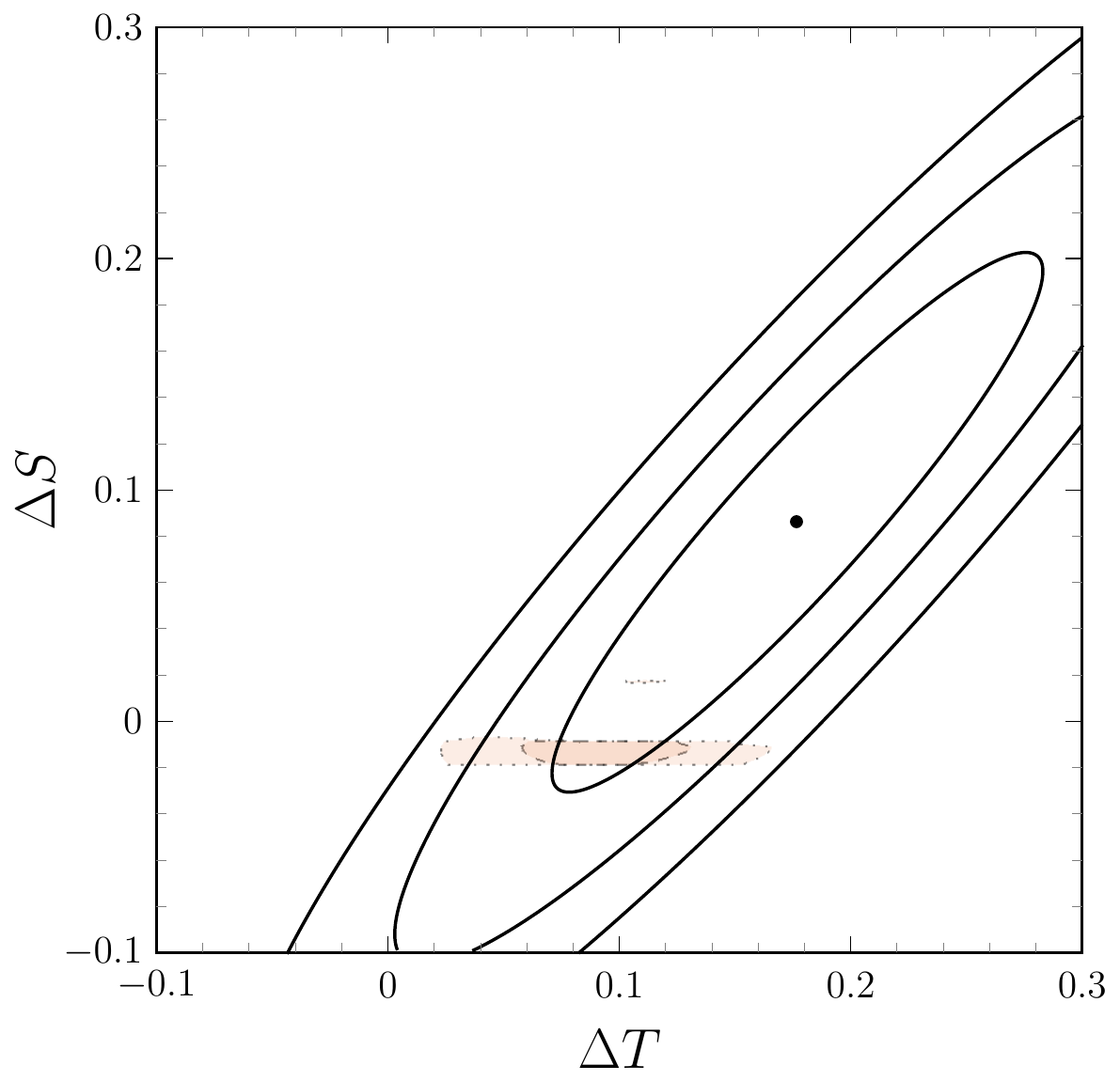}}\qquad
\subfloat[$(\Delta T,\Delta S)$ in \refeq{eq:DSDT:MW:chin}.\label{sFIG:DSDT:MW:chin}]{\includegraphics[width=0.3\textwidth]{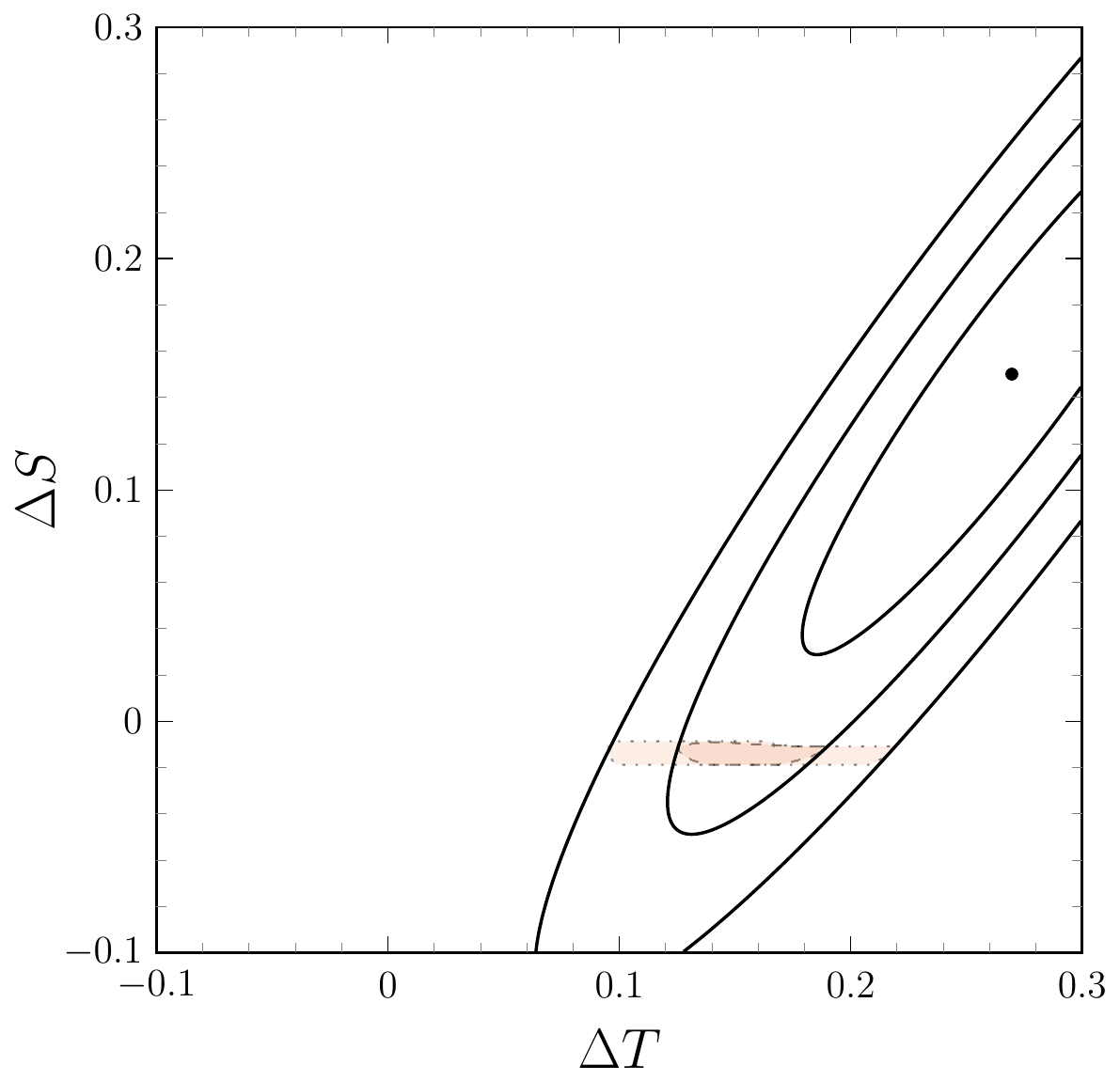}}
\caption{$\Delta S$ vs. $\Delta T$ allowed regions; the $1$, $2$ and $3\sigma$ 2D-$\Delta\chi^2$ elliptical contours from the constraints are shown.\label{FIG:DSDT:MW}}
\end{center}
\end{figure}
\section{Conclusions\label{SEC:Conclusions}}
The recent ATLAS excesses in the search of new scalar bosons decaying into a pair of $\tau$ leptons, together with the $(g-2)_{e,\mu}$ anomalies pointing towards lepton flavour universality violation, might require an extension of the SM scalar sector and a non-trivial flavour structure. Aiming to address these NP hints in a common framework, we consider a type I (or X) 2HDM with a modified lepton sector, one loop stable under renormalization, that leads to the breaking of lepton flavour universality beyond the mass proportionality. The ability of the model to solve the muon $g-2$ anomaly together with the different scenarios one can consider for the electron anomaly, related to the Cs or the Rb determination of the fine structure constant, was successfully addressed in \cite{Botella:2020xzf} and \cite{Botella:2022rte}. In this context, we further incorporate to the analyses the apparent disagreement between the observed upper limits on $\sigma(pp \rightarrow S)_{\mathrm{[ggF]}} \times \BR{S\rightarrow \tau^{+}\tau^{-}}$ and the corresponding expectation, in the invariant mass range [0.2; 0.6] TeV, presented by ATLAS. A detailed numerical analysis, taking into account all relevant low and high energy constraints, is performed. We obtain an explanation of the excess in terms of the scalar H contribution, the pseudoscalar A contribution, or both at the same time. In particular, the allowed regions in the parameter space include scalar masses $\mH \in [0.2;0.6]$ TeV and $\mA \in [0.4; 1.0]$ TeV with $\mA > \mH$, values of $\tb \in [7; 10]$ and scalar couplings to $\tau$ leptons larger than 120 GeV. In this region of the parameter space, the muon $g-2$ anomaly receives important one loop H mediated contributions, together with similar size two loop contributions coming from Barr-Zee diagrams with a top quark or a tau lepton in the closed fermion loop. The Cs electron anomaly is explained at two loops with equally important contributions from virtual top quarks and tau leptons. The main results of our discussions would not be substantially altered if one had considered other scenarios for the electron anomaly. On the other hand, the analogous ATLAS excess in $b$-associated production and the CMS excess in ditop production cannot be accommodated in this very framework. Further NP prospects concerning the anomalous magnetic moment of the $\tau$ lepton have been analysed. In particular, considering a light scalar H that couples significantly to $\tau$ leptons, a one loop dominated contribution to the $(g-2)_\tau$ anomaly of $\mathcal{O}(10^{-6})$ can be generated, which is within reach of the level of the expected sensitivity at Belle-II. Finally, the disagreement between the CDF measurement of $M_W$ and its SM prediction is translated into a change in the oblique parameters $(\Delta S, \Delta T) \neq (0,0)$. However, this has no significant impact on the regions of interest to explain ``the ATLAS excess''.
\section*{Acknowledgments}
The authors acknowledge support from Spanish \textit{Agencia Estatal de Investigaci\'on}-\textit{Ministerio de Ciencia e Innovaci\'on} (AEI-MICINN) under grants PID2019-106448GB-C33 and PID2020-113334GB-I00/AEI/10.13039/501100011033 (AEI/FEDER, UE) and from \textit{Generalitat Valenciana} under grants PROMETEO 2019-113 and CIPROM 2022-36. 
The work of FCG is funded by the Presidential Society of STEM Postdoctoral Fellowship, CWRU. 
CM is funded by \textit{Conselleria de Innovación, Universidades, Ciencia y Sociedad Digital} from \textit{Generalitat Valenciana} (grant ACIF/2021/284). 
MN is supported by the \textit{GenT Plan} from \textit{Generalitat Valenciana} under project CIDEGENT/2019/024.

\begin{appendices}
\section{Model building}\label{appendix:Model}
The general 2HDM is built as an extension of the SM where a second Higgs doublet is added. The corresponding Yukawa Lagrangian reads
\begin{equation}\label{eq:2HDMLagrYuk}
\begin{split}
\mathscr L_{\rm Y}=& -\wQLb{}\left(\SD{1}\matYukD{1}+\SD{2}\matYukD{2}\right)\wdR{} -\wQLb{}\left(\SDti{1}\matYukU{1}+\SDti{2}\matYukU{2}\right)\wuR{} \\
& -\wLLb{}\left(\SD{1}\matYukL{1}+\SD{2}\matYukL{2}\right)\wlR{}
 +\Hc\, ,
\end{split}
\end{equation}
where $\SDti{j}\equiv i\sigma_2\SDc{j}$. Electroweak spontaneous symmetry breaking is obtained with non-zero vevs
\begin{equation}
\langle 0|\SD{1}|0\rangle
=\frac{1}{\sqrt{2}}\begin{pmatrix}0\\v_1  e^{i\theta_1}\end{pmatrix}\, , \quad 
\langle 0|\SD{2}|0\rangle
=\frac{1}{\sqrt{2}}\begin{pmatrix}0\\v_2 e^{i\theta_2}\end{pmatrix}\, ,
\end{equation}
and thus the fields can be expanded around the vacuum as
\begin{equation}\label{eq:SD:expansion}
\SD{j} = e^{i\theta_j} \begin{pmatrix} \varphi_j^{+} \\ (v_j + \rho_j + i\eta_j)/\sqrt{2}\end{pmatrix}.
\end{equation}
It is always possible to rotate the scalar fields into a basis where just one of the doublets presents a non-vanishing vev. This is the so-called Higgs basis \cite{GEORGI197995,PhysRevD.19.945,PhysRevD.51.3870} defined by
\begin{equation}\label{eq:HiggsBasis:01}
\begin{pmatrix}\Hv\\ \Ho\end{pmatrix}=\HbROT\,
\begin{pmatrix}e^{-i\theta_1}\SD{1}\\ e^{-i\theta_2}\SD{2}\end{pmatrix},\quad \text{with}\quad 
\HbROT=\begin{pmatrix}\phantom{-}\cb & \sb\\ -\sb & \cb \end{pmatrix}\, ,
\end{equation}
where $\cb\equiv\cos\beta\equiv\vev{1}/\vev{}$ and $\sb\equiv \sin \beta \equiv\vev{2}/\vev{}$. The ratio among the vevs is given by $\tb\equiv\tan\beta=\vev{2}/\vev{1}$ and $\vev{}^2=\vev{1}^2+\vev{2}^2 \simeq (246\ \mathrm{GeV})^2$. In the Higgs basis, the Yukawa Lagrangian
\begin{equation}\label{eq:2HDM:YukLag:HiggsBasis}
\begin{split}
\mathscr L_{\rm Y} =&-\frac{\sqrt{2}}{v}\wQLb{}\left(\Hv\wmatMD+\Ho\wmatND\right)\wdR{} -\frac{\sqrt{2}}{v}\wQLb{}\left(\Hvti\wmatMU+\Hoti\wmatNU\right)\wuR{}\\
&-\frac{\sqrt{2}}{v}\wLLb{}\left(\Hv\wmatML+\Ho\wmatNL\right)\wlR{}+\Hc \,  
\end{split}
\end{equation}
has the simplest interpretation: $\wmatMf{}$ are the non-diagonal fermion mass matrices, coupled to the only Higgs doublet that acquires a vev, and $\wmatNf{}$ are a new set of flavour structures for the different fermions $f=d,\, u,\, \ell$.\\

In the quark sector the model is shaped by a $\ZZ$ symmetry which transforms
\begin{equation}\label{eq:Z2}
\SD{1}\mapsto \SD{1},\ \SD{2}\mapsto -\SD{2},\ \wdR{}\mapsto -\wdR{},\ \wuR{}\mapsto -\wuR{},\ \wQL{}\mapsto \wQL{}.
\end{equation}
This corresponds to a type I 2HDM (in a type II 2HDM, one would have $\wdR{}\mapsto\wdR{}$ in \refeq{eq:Z2}), and leads to $\matYukD{1}=0$, $\matYukU{1}=0$ and
\begin{equation}
\wmatND=\tbinv\wmatMD\, , \quad \wmatNU=\tbinv\wmatMU\, .
\end{equation}
On the other hand, in the lepton sector we only require that $\matYukL{1}$ and $\matYukL{2}$ are simultaneously bidiagonalized with the resulting diagonal $\wmatNL$ arbitrary. This condition cannot be enforced by a symmetry. However, these requirements on the Yukawa sector turn out to be respected by one loop RGE, and thus scalar flavour changing neutral couplings do not arise at this level.

In the Higgs basis, the expansion of the doublets around the vevs reads
\begin{equation}\label{eq:HiggsBasisDoublets}
\Hv=\begin{pmatrix} G^+\\ \frac{v+\nHH+iG^0}{\sqrt{2}}\end{pmatrix},\quad 
\Ho=\begin{pmatrix} \cHp\\ \frac{\nHR+i\nHI}{\sqrt{2}}\end{pmatrix}\, .
\end{equation}
We can readily identify the would-be Goldstone bosons $G^\pm$ and $G^0$, the new physical charged Higgs fields $\cH$ and three neutral scalars $\{\nHH,\,\nHR,\,\nHI\}$ that are rotated into the mass eigenstates by 
\begin{equation}\label{eq:PhysNeutralScalars}
\begin{pmatrix}
\nHH\\ \nHR\\ \nHI
\end{pmatrix}=\ROTmat\begin{pmatrix}\nh\\ \nH\\ \nA\end{pmatrix}\, ,
\end{equation}
where the matrix
\begin{equation}
    \ROTmat=\begin{pmatrix}\sba & -\cba & 0\\ \cba & \phantom{-}\sba & 0 \\0 & 0 & 1\end{pmatrix}\, 
\end{equation}
is block diagonal if we assume that there is no CP violation in the scalar sector. On that respect, we recall that the scalar potential is shaped by a $\ZZ$ symmetry that is softly broken. As previously mentioned, $\sba \equiv \sin(\alpha + \beta)$ and $\cba \equiv \cos(\alpha + \beta)$, being $\pi/2 - \alpha$ the mixing angle that parametrizes the change of basis from the fields in \refeq{eq:SD:expansion} to the mass eigenstates in \refeq{eq:PhysNeutralScalars}.\\ 
The scalar potential is
\begin{multline}\label{eq:scalarpot}
V(\SD{1},\SD{2})=\mu_{11}^2\SDd{1}\SD{1}+\mu_{22}^2\SDd{2}\SD{2}+\mu_{12}^2(\SDd{1}\SD{2}+\SDd{2}\SD{1})+\lambda_1(\SDd{1}\SD{1})^2+\lambda_2(\SDd{2}\SD{2})^2\\
+2\lambda_3(\SD{1}\SD{1})(\SDd{2}\SD{2})+2\lambda_4(\SDd{1}\SD{2})(\SDd{2}\SD{1})+\lambda_5((\SDd{1}\SD{2})^2+(\SDd{2}\SD{1})^2),
\end{multline}
where $\mu_{ij}^2$, $ij=11,22,12$, and $\lambda_j$, $j=1,\ldots,5$ are real. As mentioned, $V(\SD{1},\SD{2})$ is $\ZZ$ symmetric (see \refeq{eq:Z2}), except for the soft breaking term with $\mu_{12}^2$. Instead of the 8 parameters $\{\mu_{ij}^2,\lambda_j\}$ in \refeq{eq:scalarpot}, it is convenient to use $\vev{}$, $\mh$, $\mH$, $\mA$, $\mcH$, $\tb$, $\cba$ and $\mu_{12}^2$, and fix $\vev{}\simeq 246$ GeV, $\mh\simeq 125$ GeV ($\nh$ is the SM-like Higgs). The remaining free parameters of the model, necessary for the numerical analysis, are just $\nl{e}$, $\nl{\mu}$ and $\nl{\tau}$.

\end{appendices}
}
\printbibliography

\end{document}